\documentclass[twocolumn]{autart}    

\usepackage{graphicx}         
\usepackage{amsmath} 
\usepackage{amssymb}  
\usepackage{url}
\usepackage{graphicx}
\usepackage{tikz}
\usetikzlibrary{arrows,shapes,fit,calc,backgrounds}
\usepackage{circuitikz}
\usepackage{enumerate}
\usepackage{comment}
\usepackage{cite}
\usepackage{algorithm}
\usepackage{algpseudocode}
\newcounter{taskcounter}
\newenvironment{hypot}{ \refstepcounter{taskcounter}\textbf{A\arabic{taskcounter}}:}
{}

\renewcommand \b{\mathbf}			

\newcommand {\ass}[1]{\textbf{A\ref{#1}}}

\newcommand \teq{\triangleq}			
\newcommand \R{\mathbb{R}}			
\newcommand \M{\mathcal{M}}			
\newcommand \T{\mathcal{T}}			
\newcommand \probs{(\Omega,\mathcal{F},P)}	
\newcommand \U{\mathcal{U}}			
\newcommand \A{\mathscr{A}}			
\newcommand \conv{\rightarrow}			
\newcommand \D{\Delta}				
\newcommand \E{\mathbb{E}}			
\newcommand \degree{\ensuremath{^\circ}C}	

\def \ck{\b{C}_{k}[0,\infty) }
\def \cl{\b{C}_{l}[0,\infty) }
\def \f{\infty}

\def \xn{x^{(N)}}
\def \xf{\bar x}
\def \A{\mathcal A}
\def \rn{1 \leq i \leq N}
\def \rt{t \in [0,\infty)}
\def \c{\b{C}[0,\infty)}
\def \cb{\b{C}_b[0,\infty)}

\def \G{\mathcal{G}}

\usepackage{soul}
\newcommand{\ark}[1]{\textcolor{black}{#1}}
\newcommand{\rma}[1]{\textcolor{black}{#1}}
\newcommand{\rsa}[1]{\textcolor{black}{#1}}
\begin{document}

\begin{frontmatter}

\title{AN INTEGRAL CONTROL FORMULATION OF MEAN FIELD GAME BASED LARGE SCALE COORDINATION OF LOADS IN SMART GRIDS\thanksref{footnoteinfo}} 

\thanks[footnoteinfo]{This paper was not presented at any IFAC 
meeting. Corresponding author R. Salhab. Tel. +1-(514)-430-4320.}
\thanks[footnoteinfoArman]{The first author was at GERAD and \'Ecole Polytechnique de Montr\'eal when this work was carried out.}

\author[Automat]{Arman C. Kizilkale\thanksref{footnoteinfoArman}}\ead{arman@automat.ai},    
\author[Montreal,Gerad]{Rabih Salhab}\ead{rabih.salhab@polymtl.ca},               
\author[Montreal,Gerad]{Roland P. Malham\'e}\ead{roland.malhame@polymtl.ca}  

\address[Montreal]{Department of Electrical Engineering, \'Ecole Polytechnique de Montr\'eal, H3T 1J4 Montreal, Canada}

\address[Gerad]{Group For Research in Decision Analysis (GERAD), H3T 1J4 Montreal, Canada}

\address[Automat]{Automat, H2X 2T6 Montreal, Canada}
          

\begin{abstract}                   
Pressure on ancillary reserves, i.e.frequency preserving, in power systems  has significantly mounted due to the recent generalized  increase of the fraction of (highly fluctuating) wind and solar energy  sources in  grid generation mixes.  The energy storage  associated with millions of individual customer electric thermal (heating-cooling)  loads is considered as a tool for smoothing power demand/generation imbalances. The piecewise constant level tracking problem of their collective energy content is formulated as a  linear quadratic mean field game problem with integral control in the cost coefficients. The introduction of integral control brings  with it a robustness potential to mismodeling, but also the potential of cost coefficient unboundedness. A suitable Banach space is introduced to establish the existence of Nash equilibria for the corresponding infinite population game, and algorithms are proposed for reliably computing a class of desirable near Nash equilibria. Numerical simulations illustrate the flexibility and robustness of the approach.
\end{abstract}

\end{frontmatter}

\section{Introduction}\label{intro}

Since the seventies, load management via 
direct load control and its predominantly 
price induced variant, demand response, have 
been considered as tools for shaping the 
load demand in power systems so as to 
achieve peak load shaving and valley filling 
\cite{mitchell1977electricity}. Such 
measures help defer generation and capacity 
expansion, and operate generators in the 
vicinity of their most efficient operating 
point. However, the seriousness with which 
demand response (or load dispatch) are being 
considered is a relatively recent 
phenomenon. It is mainly caused by the 
increasing share of intermittent renewable 
energy sources (such as wind and solar) in 
the energy mixes of power producers 
worldwide 
\cite{calenergycom,worldwatch,nytimes}. The 
ensuing variability and diminished 
predictability of generation has indeed put strong pressures on the ability of independent system operators to maintain grid stability and insure reliable power delivery and transmission to consumers in their power pool \cite{hirth2013control,ackermann2015integrating}.
  
One of the striking characteristics of 
so-called smart grids is an ability to rely on both an increasingly pervasive communication network, and an improved electricity distribution network, to shift the responsibility of balancing electricity demand with power generation from being solely a generation side task, to one shared increasingly by both producers and customers. Furthermore, the role of consumers is  gradually changing in that they can also contribute to power generation or delivery, mostly through rooftop solar panels, electric batteries  in dwellings or electric vehicles  (EV's)\cite{el2016prospect}. \rma{ In that context, pricing or control dictated coordination of certain deferrable classes of loads (ex. pool pump loads, or energy storage capable loads such as electric space heaters or air conditioners) to compensate for fluctuations in intermittent renewable} \ark{energy} \rma{has become   increasingly attractive relative to expensive battery storage alternatives.  Dispersed energy storage for frequency regulation in the presence of wind
energy is investigated in \cite{2009Cal}. Pricing based demand response 
of large commercial buildings is analyzed in \cite{mathieu2011variability}, while state estimation for the direct control of thermostatic populations is considered in \cite{mathieu2013state}; domestic heating systems are employed as
heat buffers in \cite{2011TSAR_CDC}, while a coordinated  randomized control of Florida pool pumps is considered in \cite{meyn2015ancillary}, and the related state estimation problems are studied in \cite{chen2017state}. A decentralized mean field based
charging control strategy for large populations of plug-in
electric vehicles (PEVs) 
is presented in \cite{ma2013decentralized}.}

\rma{The resulting  drastically modified electric grid landscape has produced: (a) \textit{new modeling requirements} as power system operators will increasingly need bottom up type models to anticipate the aggregate behavior of  customer loads to real time demand response pricing signals, in particular synchronization effects  and the ensuing load rebound  (see \cite{muratori2016residential}); or the effects of direct control signals in load dispatch approaches; (b) \textit{new large scale load control challenges}.} \rma{This is particularly true when residential type load controls, or EV charging coordination are considered for demand dispatch, as the sheer number of control points (possibly in the millions) needed to achieve significant system impact make it essentially impossible to monitor centrally every load, and compound computational challenges; in that case, effective forms of hierarchical coordination with decentralized/distributed controls are most advisable (see \cite{callaway2011achieving,meyn2015ancillary}). They  allow  scalability as  computations become parallelized, provide more resilience to communication failure, an increased level of privacy,  and a guarantee that local constraints can be locally satisfied. Decentralized control is also consistent with the multi-agent framework advocated for smart grids and microgrids in \cite{mcarthur2007multi,hasanpor2017transactive}}. As a result,  on the modeling front, the approaches inspired from statistical physics, which start from physically based stochastic microscopic descriptions of loads and build via ensemble analysis macroscopic aggregate descriptions, and started in the eighties (\cite{1985MC_TAC,1994LM}), are making a strong comeback (see \cite{2009Cal,2013TLW,chen2017state,ma2013decentralized,2013ZLCK_PES}). On the control theoretic front, game theory has witnessed a surge. More importantly though, one has to note the relatively novel development of so-called mean field games (MFG's) (\cite{2006LL,2006HMC_CIS,2007HCM_TAC,Caines2017}): they combine in effect the 
\rma{bottom-up} modeling power of statistical 
physics with the decentralized control 
theoretic potential of game theory. They are 
at their most basic level a game theory of 
large groups of class wise interchangeable 
agents, whose individual influence on the 
group asymptotically vanishes with the size, 
thus leading in the limit, under Nash 
equilibrium conditions to (i) predictability 
of the group behavior via the law of large 
numbers, (ii) decentralized individual 
control policies with a dependence on local 
state and statistical information on the 
agents' dynamic and cost parameters, as well 
as their initial states distribution. 

\rma{MFG approaches have found applications in 
numerous areas including economics 
(see\cite{7526095} and the references 
therein), and engineering systems in general 
including CDMA communication systems, virus 
propagation in computer systems, crowd 
dynamics, traffic analysis, etc. (see 
\cite{ElectronEng-01-00018} for a survey)}.

In this paper, we intend to illustrate how MFG's constitute a natural tool in aggregator based load dispatch, with the particular class of thermostatic electric heating loads aimed at. An aggregator is considered as a technically equipped broker, acting on behalf of a pool of customers to coordinate their loads as a virtual battery on the energy markets. Her role is to rely on a regularly updated aggregate load model to assess the time dependent load absorption or relief potential of the load pool under its responsibility, and to compute and send piecewise constant mean pool energy content targets which are feasible under comfort and security constraints, and which will result in overall economic gains for the pool, and indirectly, the quality of the environment. A MFG based computation of individual load control strategies will lead to decentralized locally computable and implementable controls thus achieving computational scalability, with minimal communication requirements, local compliance with comfort and security constraints, and as we shall see, a size of device contribution in direct relation to the device current ability to contribute.

Besides the application context, on the technical level, the fundamental novelty here is the introduction of integral control in the cost coefficients of what is otherwise an instance of a linear quadratic Gaussian MFG \cite{2007HCM_TAC}. The integral control in the cost coefficient introduces with it both the potential benefits of robustness of integral controllers to inadequate modeling and disturbances which has been so successfully leveraged in applications \cite{ortega2002nonlinear}, and the unfortunate risk of a cost coefficient going unbounded. After setting up our microscopic controlled device model and defining a cost function which weighs the individual's reluctance to contribute against the objective of having to meet the global aggregate target, we establish the existence of an asymptotic Nash equilibrium on an appropriately defined Banach space. This equilibrium may be on target (desirable equilibrium), or may correspond to an undesirable equilibrium (a cost coefficient going unbounded).  We then develop an algorithm to reliably compute a near Nash strategy corresponding to a desirable equilibrium. Numerical results illustrate the flexibility of the approach and the robustness to modeling errors conferred by the integral control structure. 

The rest of the paper is organized as follows.  
In Section \ref{hm} we introduce the model that will be used throughout the paper and propose our collective target tracking mean field formulation. In Section \ref{fpa} we present a fixed point analysis for the equation system characterizing the limiting mean field, and we develop our $\epsilon$-Nash Theorem indicating that an approximate Nash Equilibrium is attained. We provide in Section \ref{num} an algorithm that
generates approximate desirable fixed point mean trajectories. Lastly, in Section \ref{sec:sim}, we provide simulation results together with comparisons to a prevailing target tracking control formulation.

The following notation is defined; the set of nonnegative real numbers is denoted by $\R_+$.
The set $\c$ denotes the family of all continuous functions on $[0,\infty)$, $\cl=\{x: x \in \c,\exists k_0>0,0\leq x(t)\leq k_0t, \forall t\geq 0 \}$, $\cb=\{ x:x\in \c,\sup_{t\geq 0} | x_t | < \f\}$, and for any $x\in\b{C}_b$, $\lVert \cdot \rVert_\f$ denotes the supremum norm: $\lVert x \rVert_\f\teq \sup_{t\geq 0} | x_t |$.
\section{ELECTRIC SPACE HEATER MODELS}\label{hm}

In the following, we first introduce the model for space heating dynamics that will be employed throughout the paper. 
Subsequently, a \emph{collective target tracking mean field} control model is defined together with the individual control actions and the corresponding mean field system of equations is developed.


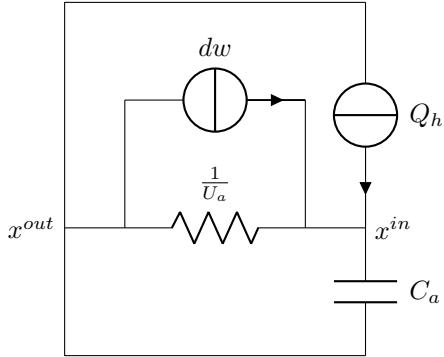
\begin{figure}[th]
\centering
\begin{circuitikz}
\draw (0, 0) node[left]{$x^{out}$}  to [R=$\frac{1}{U_a}$] (4,0) node[right]{$x^{in}$} ;
\draw [C=$C_a$] (4,0) to (4,-1.7) ;
\draw (4,-1.7) to (0,-1.7) to (0,0);
\draw [I = $Q_h$] (4,3) to (4,0);
\draw (4,3) to (0,3) to (0,0);
\draw [I = $dw$] (0.8,1.7) to (3.2,1.7);
\draw (0.8,1.7) to (0.8,0);
\draw (3.2,1.7) to (3.2,0);
\end{circuitikz}

\caption{Equivalent thermal parameter (ETP) model of a household}
\label{etp}
\end{figure}

We employ a one dimensional equivalent thermal parameter (ETP) model (see Fig.\ \ref{etp}) \cite{1978So} to describe the thermal dynamics of a single household, which is written as 
\begin{equation}\label{eqn:etp}
d x_t^{in} = \frac{1}{C_a} [ -U_a (x_t^{in}-x^{out}) + Q_h(t) ]dt + \sigma dw_t, \end{equation}
where $t\geq 0$, $x^{in}$ is the air temperature inside the household, $x^{out}$ is the outside ambient temperature, $C_a$ is the thermal mass of air inside, $U_a$ is conductance of the walls and $Q_h$ is the heat flux from the heater. Note that $w_t,\, t\geq 0,$ is a standard Wiener process defined on $\probs$ to reflect the noise on the system caused by random processes of heat gain and loss due to customer activity within the dwellings, and $\sigma$ is its volatility term.

For brevity of notation, the system for heater $\A_i,\,\rn,$ is equivalently written as
\begin{equation}\label{eqn:heatdyn}
dx_t^i=[-a^i (x_t^i - x^{out,i}) + b^i u_t^i] dt + \sigma dw_t^i,
\end{equation} 
where $x^i \teq x^{in,i}$ and $u^i \teq Q_h^i$ for $\rn$; $a^i=U_a^i/C_a^i$ and $b^i=({C_a^i})^{-1}$. 
$\{w^i,1 \leq i \leq N\}$ are $N$ independent Wiener processes on a probability space $(\Omega,\mathcal{F},\mathbb{P},\{\mathcal{F}_t\}_{t \in [0,T]})$. $\{\mathcal{F}_t\}_{t \in [0,T]}$ is the augmented filtration
of $\{\sigma(x^i_0,w^i_s,0 \leq s \leq t,1 \leq i \leq N)\}_{t\in [0,T]}$
\cite[Section 2.7]{1988KS}, where $x^i_0$, $1 \leq i \leq N$, are the initial 
temperatures assumed i.i.d. with distribution 
$P_0$ and finite 
second moments
$\mathbb{E} |x^i_0|^2 < \infty$, and also independent
of $\{w^i,1 \leq i \leq N\}$. 
Note that this model is similar to the model given in \cite{1985MC_TAC}, where the thermostat control is exchanged with a linear control.

We consider that users ``naturally" would like their devices to stay at their initial temperatures (normally attained via thermostatic action and before the intervention of the power utility control center). Thus,  we have reformulated the control effort as the signal required to make them deviate from that initial  temperature; more precisely, the control effort to stay at the initial temperature is considered free and will not be penalized by the cost function to be defined. The corresponding dynamical equation is written as 
\begin{equation}\label{eqn:heatdynU}
dx_t^i=[-a^i (x_t^i - x^{out,i}) + b^i (u_t^i+u^{free,i})] dt + \sigma dw_t^i,
\end{equation}
where $u^{free,i} \teq {(b^i)}^{-1}a^i(x_0^i - x^{out,i})$.

\begin{hypot}\label{x0Ass}
\rsa{We assume that the  mean temperature is bounded from above and below by comfort levels; i.e., $l \leq \E x_0^i \leq h$.}\\ \rma{Note that $l$ and $h$ respectively represent the lowest  and highest  temperatures considered tolerable in the controlled dwellings (see Fig.3 below). These numbers are to be initially agreed upon with the participants in the control program (In our simulations, we took them to be respectively 17 and 25 degrees celsius).}\\
We assume a population of $m$
types of agents, that is, the vector of individual 
parameters $\theta^i:=(a^i,b^i,x^{out,i})$ takes values in 
a finite set  $\{\Theta_1,\dots,\Theta_m\}$, which does not depend on the size 
of the population $N$.   
The empirical probability measure of the sequence $\{\theta^i\}_{i=1,\dots,N}$ is denoted by     
$P^N_\theta(\Theta_s) = 1/N 
\sum_{i=1}^N 1_{\{\theta^i = \Theta_s
\}}$ for 
$s=1,\dots,m$. 
We assume that $(P^N_\theta(\Theta_1),\dots,P^N_\theta(\Theta_m))$ converges to $P_\theta=(n_1,\dots,n_m)$, as $N \to \infty$, where $n_s>0$ for all $1 \leq s \leq m$. 
\end{hypot} 
\subsection{
\rma{General Characteristics of a Desirable Load Coordination Scheme }\label{sec:Charac}}
\rma{In this subsection, we briefly discuss the important characteristics of a desirable load coordination scheme by the aggregator. Given the potentially very large number of residential loads  that the aggregator needs to coordinate to form  a virtual battery of sufficient size, it is impossible to assume it will be able to centrally observe all loads. However, the aggregator can be assumed to have the ability to broadcast common signals such as a temperature target $y$. Thus, a first essential characteristic of the control scheme must be \textbf{(i) Decentralization}. This is because in the absence of local state observation and given the uncertain/stochastic nature of individual loads, open loop control signals could send regularly individual temperatures  at undesirable levels, thus potentially causing customer withdrawals from the aggregator's power pool. \textit{A local controller can always use the local information to insure that comfort/security constraints are not violated}. Another important feature is that the resulting control actions should lead quickly to the desired result with least disturbance to the customer, again to avoid potential customer withdrawals. This can be summarized as \textbf{(ii) Minimality of forced temperature excursions}. A third feature which is desirable but not essential is  the minimization of the signalling overhead to achieve the objective, and summarized as \textbf{(iii) Parsimony of communications.} In the following, we discuss the advantages, yet shortcomings  of  a simple, intuitive, standard LQG tracking  controller which would have every device track the desired mean temperature level $y$, and later compare its would be  performance to that of our proposed MFG Based integral controller. }

\subsection{
\rsa{Advantages and Shortcomings of a }Standard LQG Tracking Controller}\label{sec:clasTrack}

\rma{Assuming the aggregator  has set the objective of attaining a given population mean  temperature target $y$, a simple  and effective control scheme for doing so is to broadcast $y$ to all devices, and have each solve an LQG  based control scheme for tracking level $y$. For example, the individual cost function of devices would be as follows:} 
\begin{equation}\label{eqn:clasLQGcost}
J_i^{LQ}(u^i) = \E \int_0^\f e^{-\delta t}
\left [ ( x_t^i - y )^2{q}^{LQ} +  (u_t^i)^2 r \right ] dt.
\end{equation}
\rma{Clearly, such a control scheme, besides being very simple to implement,  meets the criteria \textbf{(i)} and \textbf{(iii)} above, i.e. it is both decentralized and communications parsimonious. However, the fundamental shortcoming of such a controller is that, while the aggregator is interested only in having the mean population state settle at target $y$, this controller sends instead \textit{all devices} to that target. This not only maximizes individual user temperature excursions, thus falling short on criterion \textbf{(ii)} above, but furthermore will result in general in a slow controller since some devices are "undoing" the contribution of other devices (some devices are heating further, while others are cooling in order to reach a mean state $y$.) }
 
\rsa{In the coordination scheme we propose next, instead of each agent state trying to track a common  target $y$, the individual cost structures (see \eqref{eqn:modCost} below) are formulated such that ultimately, it is only \textit{the mean} of the population trajectories that tracks the desired target. In the resulting MFG, the novelty is that the mean field effect is mediated by the \emph{quadratic cost function parameters} under the form of an integral error. The resulting concept will be called \textit{collective  mean target tracking}.}

\subsection{The Collective Mean Target Tracking Model}\label{sec:cttmfm}


We employ the dynamics for the heaters given in \eqref{eqn:heatdyn}, \rma{with the input redefined based on \eqref{eqn:heatdynU} }. \rma {Following  a prescriptive game theoretic view of the coordination problem,} the infinite horizon discounted cost function for agent $\A_i,\rn$, is defined as:
\begin{align}\label{eqn:modCost}
J_i(u^i,u^{-i}) = &\E \int_0^\f e^{-\delta t}
\Big [ \frac{q^{y}_t}{2}( x_t^i - z )^2 \nonumber \\
&+ \frac{q^{x_0}}{2}( x_t^i - x_0^i )^2  + \frac{r}{2} (u_t^i)^2  \Big ] dt.
\end{align}
\rma{ We now  discuss the role of each term in the integrand in \eqref{eqn:modCost}, proceeding from left to right. $z$ is actually either one of the limiting temperatures $l$ and $h$ referred to in Assumption  \ass{x0Ass}.   $z$  is set to  $l$  if the objective is overall dwellings power/ energy decrease, or $h$ if instead it is overall  increase. The value of $z$ will thus define the direction  (up or down) according to which individual dwelling temperatures will drift. However, the pressure to move in that direction is dictated by the coefficient $q^y_t$, defined below as the integral of an increasing function of the deviation between the average state of agents and a temperature target $y$.}  \rma{Thus: 
\begin{equation}\label{q:eqn}
q^y_t = \left\lvert \int_0^t g(\xn_\tau - y) d\tau \right\rvert,\quad t\geq 0,
\end{equation}
where $g$ is a real increasing continuous function on $\mathbb{R}$ with $|g(x)| \leq C(1+ |x|)$ for some $C>0$, $\forall x\in \mathbb{R}$, $y$ is the main control center dictated mean target constant level, and $\xn \triangleq (1/N) \sum_{i=1}^N x^i$.
Note that the time varying coefficient $q^y_t$ is \textit{the only channel through which the mean field} (mean dwellings temperature in our case) \textit{influence is felt by individual agents}.\\
The target $y$ is decided upon by the aggregator. The latter uses a macroscopic aggregate model of the dwellings to evaluate their current energy storage, and their short term power/ energy release  or increase potential so as to make bids on the energy market. Alternatively, in a microgrid for example, the aggregator  could also be buiding an optimal dwellings energy content schedule, based  on forecasts of the intermittent renewable power that will be available in the microgrid. We further note that energy content targets are equivalent to mean temperature targets.\\  While the first term was introduced to mathematically capture the aggregator's objective of producing upwards/downwards drift of average dwellings temperature, the next term will simulate resistance of individuals to changes in their initial temperature at the start of the control period. Indeed, the associated cost increases whenever $x^i_t$ moves away from  its initial value $x_0^i$. Note that this introduces a further element of heterogeneity amongst agents. Finally, the last term penalizes the control effort, again  redefined (recall  \eqref{eqn:heatdynU})  as the agent effort involved in moving away from $x_0^i$. In summary, while  the first term  favors the aggregator,  the next two terms in the cost integrand favor the individual agents. } 

The justification for the above cost function is that by pointing individual agents towards what is considered as the minimum (or maximum) comfort temperature $z$, it dictates a global decrease (or increase) in their individual temperatures. This pressure persists as long as the differential between the mean temperature and the mean target $y$ is high. The role of the integral controller is to mechanically compute  the \emph{right} level of penalty coefficient $q^y_t,\rt,$ which, in the steady-state, should maintain the mean population temperature at $y$. When this happens, individual agents reach themselves their steady states (in general different from $y$ and closer to their initial diversified states than standard LQG tracking would dictate). Furthermore, if the  \rsa{mean state of }agents is observed, this integral control adds robustness to model imperfections or inaccurate ambient temperature estimation. \rma{ Note that other approaches to robustness in a MFG context are possible \cite{huang2017robust,wang2017social}}. \rma{ However, they could not produce exact tracking of targets as integral control based algorithms can. }

 For each heater $i$, the set of admissible
control laws is the set \rsa{$\mathcal{U}_i$} of $\mathcal{F}_t-$progressively measurable 
\rsa{$u^i_t$} such that \rsa{$\mathbb{E}\int_0^\infty e^{-\frac{2\delta}{3}t}|u_t^i|^2dt<\infty$}. It should be noted that whenever \rsa{$(u^i,u^{-i}) \in \prod_{i=1}^N\mathcal{U}_i$}, then $J_i(u^i,u^{-i})<\infty$. 
The $2\delta/3$ in the definition of \rsa{$\mathcal{U}_i$} is to force the term 
with $q_t^y(x_t^i-z)^2$ in the cost to be finite.

In order to derive the limiting infinite population MF equation system, we start this time assuming a given (albeit initially unknown) cost penalty trajectory $q^y \in \cl$ and constant $q^{x_0}$. We show later in Lemma \ref{lem_3} that the candidates $q^y$  
 actually belong to $\cl$. Given $q^y$ and $q^{x_0}$, individual agents $\A_i,\, 1\leq i \leq N$, solve a standard target tracking LQG problem \cite{1989AM} with time varying cost coefficient. Using  techniques similar to
those used in \cite[Lemma A.2]{2010Hu_SIAM}, one can show that the optimal control law
of this LQG problem is,
\begin{equation}\label{eqn:colOpt} 
(u^i_t)^\circ=-b^i r^{-1}  (\pi^i_t x^i_t + \alpha^i_t - \pi^i_t z), \quad t\geq 0,
\end{equation}
with $\pi^i$ and $\alpha^i$ respectively the unique solution and unique bounded solution of
\begin{align} 
\label{colPi:eqn}&\frac{d\pi_t^i}{dt} = (2a^i+\delta)\pi_t^i + \frac{{(b^i)}^2}{r} (\pi_t^i)^2  - q^y_t - q^{x_0},\\
\label{colS:eqn}& \frac{d\alpha^i_t}{dt} = (a^i + \delta + \frac{{(b^i)}^2}{r} \pi_t^i)  \alpha^i_t - (a^i \pi_t^i - q^{x_0})(x_0^i-z).
\end{align}
The calculation of the unknown $q^y_t,\, t\geq 0,$ is obtained by requiring that $q^y_t$ be such that when individual agents implement their associated best responses, they  collectively replicate the posited $q^y_t$, trajectory. This fixed point requirement leads to the specification below of the collective target tracking MF equation system. In the remainder of this paper
a superscript $s$ refers to a heater of type $s$ . 
\begin{defn}
\emph{Collective Target Tracking (CTT) MF Equation System \rma{refers to the following coupled system of integro-differential equations on $\rt$}:}
\begin{align}
 q^y_t &= \left\lvert \int_0^t g(\bar x_\tau - y) d\tau \right\rvert, \label{eqn:qwatMFgen}\\
\frac{d\pi^s_t}{dt} &= (2a^s +\delta)\pi^s_t + \frac{(b^s)^2}{r}  ({\pi^s_t})^2 - q^y_t - q^{x_0}  \label{eqn:piwatMFgen}\\
\frac{d\alpha^s_t}{dt} &= (a^s + \delta + \frac{({b^s})^2}{r} \pi^s_t)  \alpha^s_t - (a^s \pi^s_t - q^{x_0})(\bar x_0 - z)\label{eqn:swatMFgen}\\
\frac{d\bar x^s_t}{dt} & = -(a^s +\frac{({b^s})^2}{r} \pi^s_t ) \bar x^s_t -\frac{({b^s})^2}{r}  (\alpha^s_t -\pi^s_t z)+ a^s \bar x_0 \label{eqn:barwatMFgen}\\
\bar x_t & = \sum_{s=1}^m n_s \bar x_t^{s},\label{eqn:meanwatMFgen}
\end{align}
\end{defn}
$1\leq s \leq m$, \rsa{where $n_s$ is defined below Assumption A\ref{x0Ass}}.

\rsa{Equation \eqref{eqn:piwatMFgen} is the Riccati equation that corresponds to the LQG optimal tracking problem \rma{solved by an} agent of type $s$.}The \rma{solution} of the collective target tracking (CTT) MF equation system \eqref{eqn:qwatMFgen}-\eqref{eqn:meanwatMFgen} is obtained offline locally by each agent only based on statistical information $P_\theta=(n_1,\dots,n_m)$ and $P_0$, \rma{assumed } available at the start of the control horizon. In theory, the control scheme is fully decentralized; i.e., no communication \rma{needs to take} place among the controllers throughout the horizon. In practice however, because of the anticipated prediction error accumulations over time, \rma{it is more advisable} to readjust periodically the control laws over long intervals based on aggregate \rma{ system measurements based on a limited set of randomly sampled agents} (see \rsa{Fig.} \ref{ca}). \rma{As we shall see in Section \ref{sec:sim}, this also helps robustifying the control performance against mismodeling}.

Note that the MF Equations for this model are significantly different from (4.6)-(4.9) in \cite{2007HCM_TAC} which is  amenable to analysis within a linear systems framework,  while uniqueness of the fixed point is obtained via a  sufficient contraction condition. In contrast, system \eqref{eqn:qwatMFgen}-\eqref{eqn:meanwatMFgen} is fundamentally nonlinear (because of the form of $q^y_t$), and \rma{$q^y_t$} itself could indeed   become unbounded. Special arguments have to be developed for the analysis.

\tikzstyle{block} = [draw, rectangle, minimum height=3em, minimum width=6em]
\tikzstyle{input} = [minimum height=3em, minimum width=6em]
\tikzstyle{output} = [minimum height=.01em, minimum width=.01em]

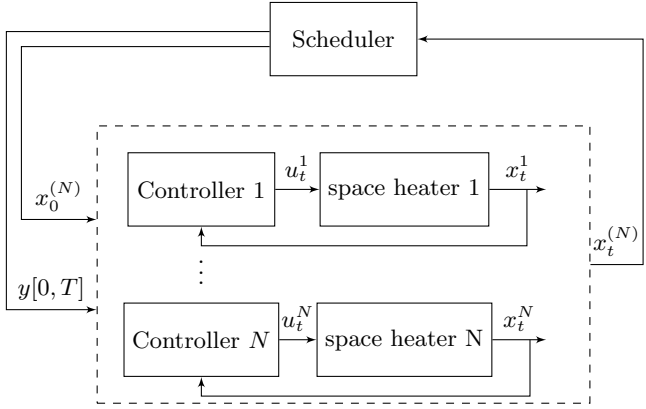
\begin{figure}
\centering
\begin{tikzpicture}[auto, node distance=4cm,>=latex']
 
\footnotesize
\node [block] (c1) {Controller 1};
\node [input, below of = c1, node distance=1cm] (c2) { \vdots };
\node [block, below of = c2, node distance=1cm] (cN) {Controller $N$};

\node [block, right of = c1, node distance=2.7cm] (w1) {space heater 1};
\node [block, right of = cN, node distance=2.7cm] (wN) {space heater N};

\node [output, right of = w1, node distance=2cm] (o1) {};
\node [output, right of = wN, node distance=2cm] (oN) {};

\node [draw,dashed,inner sep=.35cm,rectangle,fit=(c1) (c2) (cN) (w1) (wN) (o1) (oN)] (pop) {};

\node [block, above of = pop, node distance = 3cm] (s) {Scheduler};

\path (c1) edge[->] node {$u^1_t$} (w1);
\path (cN) edge[->] node {$u^N_t$} (wN);
\path (w1) edge[->] node {$x^1_t$} (o1);
\path (wN) edge[->] node {$x^N_t$} (oN);

\draw [->] ($(w1.east) + (.5cm,0)$ ) -- ++(0,-.75) -| (c1.south); 
\draw [->] ($(wN.east) + (.5cm,0)$ ) -- ++(0,-.75) -| (cN.south); 
\draw [->] ($(s.west)+(0,-.1cm)$) -| ($(pop.west) + (-1.0cm,.6cm)$) -- node {$\xn_0$} ($(pop.west) + (0,.6cm)$) ;
\draw [->] ($(s.west)+(0,.1cm)$) -| ($(pop.west) + (-1.2cm,-.6cm)$) -- node {$y[0,T]$} ($(pop.west) + (0,-.6cm)$) ;
\draw [->] (pop.east) -- node {$ \xn_t$} ($(pop.east) + (0.7cm,0)$) |- (s.east);

\end{tikzpicture}

\caption{Control Architecture in Practice}
\label{ca}

\end{figure}
\section{FIXED POINT ANALYSIS}\label{fpa}
 
\begin{figure}[th]
\centering
\includegraphics[width=.7\hsize]{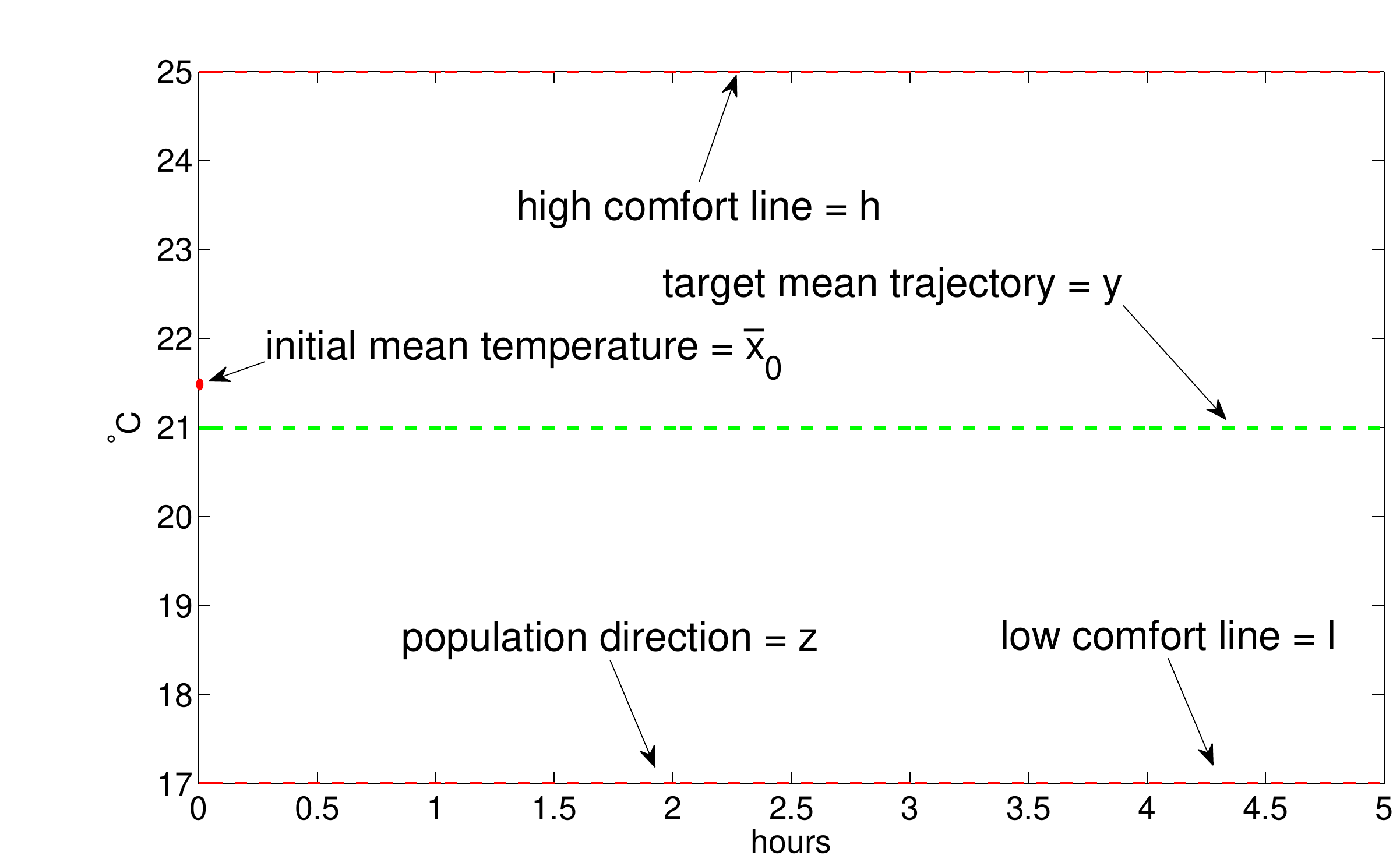}
\caption{An Energy Release Situation $(z\leq y \leq \bar x_0)$}
\label{legend}
\end{figure}

We now tackle the fixed point analysis for \eqref{eqn:qwatMFgen}-\eqref{eqn:meanwatMFgen}.
We consider the case $y\leq \bar x_0$, i.e. when the target temperature \rma{set by the aggregator} is less than or equal to the initial mean temperature of the population. The analysis is very similar for the case $\bar x_0\leq y$; which therefore will be omitted. Note that for $y \leq \bar x_0$,  $z$ is set to be less than $y$; i.e., $z=l \leq y \leq \bar x_0 \leq h$, so that the agents \emph{collectively decrease} their temperatures by moving towards that target (see \rsa{Fig.} \ref{legend}).

\subsection{CTT MF Equation System}\label{cctmf}

Considering the CTT MF Equation System \eqref{eqn:qwatMFgen}-\eqref{eqn:meanwatMFgen}, we introduce in the following the operators $\Delta$ and $\T^s$, $1 \leq s \leq m$, where $\M\teq 
\sum_{s=1}^m n_s\T^s \circ \Delta,\, \M(\bar x):\cb\rightarrow\c$, characterizes the complete equation system . 
We define the operator $\Delta:\cb\rightarrow \cl$:
\rma{\begin{equation}
q^y_t = \left\lvert \int_0^t g(\bar x_\tau - y) d\tau \right\rvert\teq \Delta{ (\bar{x}_{\tau};  {\tau}\in [0; \f)) }(t).\label{deltaDef}
\end{equation}}
\rma{Hereon, for notational convenience, the value at time $t$ of the  function  to the extreme right-hand side of \eqref{deltaDef}, will be denoted $\Delta(\bar x)(t)$ .}
Next for each $1 \leq s \leq m$, we define $\T^s:\cl\conv\c$ for the equation system \eqref{eqn:piwatMFgen}-\eqref{eqn:barwatMFgen} with input $q^y_t$ and output $\xf_t^{s}$,
which is equivalent to
\begin{equation}\label{tDef}
\bar x_t^{s} \teq (\T^s q)(t).
\end{equation}
 
Hence, one can write the MF equation system for CTT as
\begin{align}\label{mDef}
\bar x_t = \left(\sum_{s=1}^m n_s\T^s \circ \D \right) (\bar x)(t)
 \teq( \M \bar x)(t).
\end{align}
\begin{rem}\label{rem:optinterp}
It should be noted that the image by $\T^s$ of $q$, $\T^s(q)$, is the optimal state 
of the following deterministic LQR tracking optimal control problem,
\begin{align} \label{optimalinterp}
J(u)=&\int_0^\infty e^{-\delta t}\Big(\frac{q_t}{2} (x_t - z)^2
+ \frac{q^{x_0}}{2}(x_t-\bar x_0)^2 + \frac{r}{2}u^2\Big) dt \nonumber\\
\text{s.t. } &\frac{d}{dt}x = -a^{s}(x-\bar x_0) + b^{s} u, \qquad x(0)=\bar x_0.
\end{align}
\end{rem}

We need to restrict the operator $\mathcal{M}$  on an invariant subset of a Banach Space over which it is continuous. This allows us to apply a fixed point theorem, such as Schauder's theorem \cite{1973Ru}, and show the existence of a solution 
to the CTT MF equations. As shown later, the image of $\mathcal{M}$ is included in the set $\mathcal{G}:= \{x: x \in \cb , \quad z \leq x(t) \leq x_0 \}$. Hence,
one of the candidate subsets is $\mathcal{G}\subset(\cb,\|.\|_\infty)$. Indeed, the Banach space $(\cb,\|.\|_\infty)$ was used in 
the classical LQG MFG theory. We give however in the following subsection, an example where the continuity of $\mathcal{M}$
fails to hold  on $\mathcal{G}\subset(\cb,\|.\|_\infty)$. As a result, existence of a fixed point is not guaranteed. Instead here, we propose a suitable 
norm  that weakens the topology and enlarges the space in order to force continuity.
This paves the way in Subsection \ref{sec:fpt} for our proof of the existence of a fixed point.

\subsection{Failure of continuity under the sup-norm and suitable Banach space} \label{failurecont}

We start by giving an explicit formula for
the operator $\T^s \circ \Delta$. Solving equation \eqref{eqn:swatMFgen}, and replacing the solution in
\eqref{eqn:barwatMFgen} give for all $\bar x \in \mathcal{G}$,
\begin{align}
&\left(\T^s\circ \Delta \right) (\bar{x})(t)=
\phi^s(t,0)\bar{x}_0 
 +\frac{(b^s)^2}{r}  z \int_0^t
\phi^s(t,\eta) \pi_\eta^s d\eta \nonumber\\
& +a^s \bar{x}_0 \int_0^t
\phi^s(t,\eta) d\eta+\frac{(b^s)^2}{r}  (\bar{x}_0-z) \times \label{phidef}\\
&\int_0^t
\phi^s(t,\eta) 
 \int_\eta^\infty \psi^s(\tau,\eta)
(-a^s \pi_\tau^s+ q^{x_0}) d\tau d\eta, \nonumber
\end{align}
with $\phi^s(t,\eta)=\exp\left(-\int_\eta^t(a^s+(b^s)^2\pi_\tau^s r^{-1})d\tau\right)$ and 
$\psi^s(t,\eta)=\exp\left(-\int_\eta^t(a^s+\delta+(b^s)^2\pi_\tau^s r^{-1})d\tau\right)$.
Using the fact that $(b^s)^2 \pi_t^s r^{-1}$ is 
the derivative w.r.t. $t$ of $\int_\tau^t (b^s)^2 \pi_\eta^s r^{-1} d\tau $, one can show that 
\begin{equation}\label{op_M}
\begin{aligned} 
&\left(\T^s\circ \Delta \right) (\bar{x})(t)=z+
\phi^s(t,0)(\bar{x}_0-z) \\
&
+C^s\int_0^t
\phi^s(t,\eta) \int_\eta^\infty \psi^s(\tau,\eta) d\tau d\eta. 
\end{aligned}
\end{equation}
where $C^s=((b^s)^2 r^{-1}q^{x_0} +(a^s)^2 + a^s\delta)(\bar x_0-z)$.
The proof of the following lemma 
is given in Appendix \ref{gLemmas}.
\begin{lem}\label{counterexample}
Consider $g(x)=x$, $\bar{x}_0=1$, $y=0$ and $z=-1$. The operator $\mathcal{M} : \mathcal{G} \subset (\cb,\|\|_\f) \to (\cb,\|\|_\f)$ is not continuous.
\end{lem}
In the proof of Lemma \ref{counterexample} it is shown that the continuity
of $\mathcal{M}$ does not hold because 
of the way the sup-norm evaluates the 
magnitude of a function. Indeed, the proof involves a
 sequence $\mathcal{M}(\bar x^n)$  
which converges uniformly to $\mathcal{M}(y)$
when restricted to any \textit{finite}  time horizon. But, since 
the sup-norm assigns a non-negligible value
to any function with a non-negligible 
value at the tail, $\mathcal{M}(\bar x^n)$  
does not converge to $\mathcal{M}(y)$ under this norm. Thus, we need a norm that suppresses
the value of a function at the tail. 
A good candidate is the norm $\|x\|_k:=\sup\limits_{t\in [0,\infty)}|e^{-kt}x_t|$. We also need to enlarge the 
space $\cb$ to get a Banach space under 
this new norm. Thus, we define the 
following function space,
\[
\ck=\{x: x \in \c | \sup\limits_{t\in [0,\infty)}|e^{-kt}x_t|<\infty  \},
\]
where $k>0$.
We establish in Lemma \ref{lem_1} of Appendix \ref{gLemmas} some 
properties of the spaces $\ck$, which
are used in the remainder of this paper.

\subsection{Fixed Point Theorem}\label{sec:fpt}

Following the preliminary results, we present our fixed point existence theorem, \rsa{where the proof is given in Appendix \ref{proofs}.}
We make the following technical assumption.

\begin{hypot}\label{liptass}
We assume that the function $g$ is Lipschitz
continuous on compact subsets.
\end{hypot}

\begin{thm}\label{thmFixed}
Fix $0< k <\min\limits_s a^s+\delta$. Under \ass{liptass}, the following statements hold:
\begin{enumerate}[i.]
\item for all $\bar x'$ and $\bar x''$ in
$\mathcal{G}$,
$\|\mathcal{M}(\bar x')-\mathcal{M}(\bar x'')\|_k \leq 
\lambda R_k \|\bar x'-\bar x''\|_k$, where $R_k$ is a 
positive constant that depends on $k$, and 
$\lambda$ is the Lipschitz constant of $g$
on $[-|\bar x_0-y|,|\bar x_0-y|]$.
\item the map $\mathcal{M}:\mathcal{G} \subset (\ck,\|.\|_k) \to \mathcal{G} \subset (\ck,\|.\|_k)$ has at least one fixed point.
\end{enumerate}
\end{thm}
\begin{rem}
If $\lambda$ is small enough, then 
the operator $\mathcal{M}$ is a contraction, and there exists a \textit{unique} fixed 
point $\bar x \in \mathcal{G}$.
The existence of a fixed point 
for \eqref{eqn:qwatMFgen}-\eqref{eqn:meanwatMFgen} in essence 
implies the existence of a Nash 
equilibrium for an infinite 
population game. \rma{At} the 
equilibrium, the prescribed 
control actions are the best 
responses for infinitesimal 
agents, and there is no 
unilateral profitable deviation.
\end{rem}

Finally, we establish that the infinite population Nash equilibrium of Theorem \ref{thmFixed} is in fact an $\epsilon$-Nash equilibrium when applied to the practical case of a large but finite population. The proof of this result is similar to the proof of Theorem 5.6 in \cite{2007HCM_TAC}, and is therefore omitted.
\begin{thm}\label{thm:watMain_MF_thm} \emph{CTT MF Stochastic Control Theorem:}
Assume that \ass{x0Ass} and \ass{liptass} hold. Then, the
set $\{ (u^i)^\circ;1\leq N < \infty\}$, where $(u^i)^\circ$ is defined in  \eqref{eqn:colOpt} for a fixed point $\bar x$ of $\M$, yields an $\epsilon$-Nash equilibrium in the sense that, for all $\epsilon>0$, there exists $N(\epsilon)$ such that for all $N \geq N(\epsilon)$
\begin{align*} 
J_i\left((u^i)^\circ, (u^{-i})^\circ\right)-\epsilon \leq  \inf_{u^i \in\rsa{\U_i} }& J_i\left(u^i, (u^{-i})^\circ\right).
\end{align*}
\end{thm}

\section{DESIRABLE NEAR FIXED POINT ALGORITHM}\label{num}

Theorem \ref{thmFixed} states that there exists at least one fixed
point of $\mathcal{M}$. However, it does not
guarantee the existence of  of a practically useful fixed point. A 
fixed point which is desirable in practice is a sustainable  mean trajectory (i.e.replicated as the mean of best responses to the associated pressure field), 
such that its steady-state value
is \textit{equal} to the target temperature
$y$. In the following, we propose an
algorithm that generates 
desirable approximate fixed points in case  (i) $g(x)=\mu x $, and (ii) $g$ is a continuous function equal to $\mu(e^{\beta x}-1)$ when restricted to $[z-y,\bar x_0 -y]$, and constant outside this interval. Here $\beta$ and $\mu$ are positive scalars.
In both cases, the function $g$ satisfies $|g(x)| \leq C(1+|x|)$, 
for some $C>0$, under Assumption \ass{liptass}. 
In the second case, the part of $g$ that is actually involved in determining the fixed points is $\mu(e^{\beta x}-1)$, since a fixed point always belongs to the set $\mathcal{G}$. The extension of $g$ by a constant function outside the interval $[z-y,\bar x_0 -y]$ is to force the
inequality $|g(x)| \leq C(1+|x|)$ to hold everywhere. 
In the following, we denote the 
operators $\mathcal{M}$ and $\Delta$ by $\mathcal{M}_\mu$ and $\Delta_\mu$ to emphasize their dependence on $\mu$. 
We consider only the uniform case, i.e. a population with only one type of space 
heater. Thus, we omit the superscript 
$s$ in this section. 

The idea of the algorithm is 
to construct a family of mean trajectories indexed by $\mu$, $\{\bar x(\mu)\}_\mu$, such that 
$\lim\limits_{t \to \f} \bar x_t(\mu)=\lim\limits_{t \to \f} \mathcal{M}_\mu\big(\bar x(\mu)\big)_t=y$. Subsequently,
we look for  a $\mu_*$ that minimizes 
$\|\bar x(\mu)-\mathcal{M}_\mu\big(\bar x(\mu)\big)\|_{L_2}$. \rma{As a result,}
the mean trajectory $\bar x(\mu_*)$ 
is approximately a fixed point, i.e. a near Nash equilibrium of the infinite population limit. Moreover, when the infinite population optimally
responds to this mean, its mean $\mathcal{M}_{\mu_*}\big(\bar x(\mu_*)\big)$ is \rma{ guaranteed to converge to $y$} as
$t \to \f$. In order to get $\lim\limits_{t \to \f} \mathcal{M}_\mu\big(\bar x(\mu)\big)_t=y$, the 
cost coefficient trajectory $q_t^y (\mu)= \Delta_\mu(\bar x(\mu))$ must converge 
to ${q_\infty^y}^*$, satisfying the following algebraic equations:
\begin{align*}
0&= (-2a - \delta)\pi_\f - b^2 r^{-1} \pi_\f^2  +{q^y_\f}^* + q^{x_0},\\
0 &= (-a - \delta - b^2 \pi_\f r^{-1})  s_\f + (a  \pi_\f - q^{x_0})( \bar x_0 - z),\\
0 & = (-a -b^2 \pi_\f r^{-1} ) y -b^2 r^{-1}  (s_\f - \pi_\f z) + a \bar x_0 ,
\end{align*}
where $\pi_\f,\, s_\f$ denote the steady state values for Riccati and offset equations respectively. Solving the equation system for ${q^y_\infty}^*$ yields the unique solution:
\begin{equation}\label{eqn:qInf} 
{q^y_\infty}^* = \frac{ [a (a+\delta) r + q^{x_0} b^2] }{b^2}\left ( \frac{\bar x_0 - y}{y - z} \right).
\end{equation}

Thus, it remains to find a family of 
mean trajectories $\{\bar x(\mu)\}_\mu$ that 
converge to $y$ as $t \to \f$, and such that their corresponding cost coefficient trajectories
converge to ${q^y_\f}^*$ as $t \to \f$.  Indeed, the near Nash desirable mean trajectory will be sought for within that family. 
We construct the family as follows. We 
start by applying the cost coefficient 
trajectory $q_t^y=n_1 {q_\f^y}^*$ for  $t\in [0,t_0]$ and $q_t^y={q_\f^y}^*$ for $t>t_0$, for some $n_1>1$ and $t_0>0$. This 
cost coefficient trajectory generates 
a mean trajectory $\bar x^{sup}=\mathcal{T}(q_t^y)$ which will constitute an upper boundary of our search set. Clearly, $\bar x^{sup}$ converges
to $y$, in fact exponentially so, since the $q_t^y$ that generates it becomes constant after finite time $t_0$. Similarly, we get $\bar x^{inf}$, the lower boundary of the search set, by 
applying the same cost coefficient where
we replace $n_1$ by a larger number $n_2$. 
$\bar x^{sup}$ is greater than $\bar x^{inf}$ since it is 
generated by a cost coefficient lower than that of $\bar x^{inf}$, thus corresponding to a lower pressure field \cite{Bitmead1991}. 
Afterward, we define $\mu_{sup}={q_\f^y}^*/\lim\limits_{t \to \f}\Delta_1(\bar x^{sup})$, with the integral in the denominator guaranteed to remain finite, because of the exponential convergence of the mean to the target. Thus, by construction, 
$q^{sup}=\Delta_{\mu_{sup}}(\bar x^{sup})$ converges to ${q_\f^y}^*$. 
Similarly, we define $\mu_{inf}$, such that $q^{inf}=\Delta_{\mu_{inf}}(\bar x^{inf})$ converges to ${q_\f^y}^*$. 
If we choose 
$t_0$ small, $n_1$ and $n_2$ close to one, then $\bar x^{inf}$ and 
$\bar x^{sup}$ undershoot the target temperature $y$ for a small interval of time and climb back to $y$. As a result,
the negative parts of the integrals $\int_0
^\f g(\bar x^{inf} -y )dt$ and $\int_0
^\f g(\bar x^{sup} -y )dt$ do not outweigh 
the positive parts. In this way, for
any $\mu>0$, if $\bar x^{inf}\leq \bar x \leq \bar x' \leq  \bar x^{sup}$, then $\lim\limits_{t\to \f}\Delta_\mu (\bar x) \leq \lim\limits_{t\to \f}\Delta_\mu (\bar x') $. In particular,
$\mu_{sup} < \mu_{inf}$.
Hence, for each $\mu_{sup}<\mu<\mu_{inf}$, 
$\Delta_\mu(\bar x^{sup})$ converges 
to a value greater than ${q_\f^y}^*$, while
$\Delta_\mu(\bar x^{inf})$ to a value 
lower than ${q_\f^y}^*$. We define for each $f\in [0,1]$ the trajectory $\bar x^f = (1-f) \bar x^{inf}+f\bar x^{sup}$.  Using the continuity of $\lim\limits_{t\to \f}\Delta_\mu (\bar x^f)$ with respect to $f$, one can use the dichotomy method to find  for each $\mu_{sup}<\mu<\mu_{inf}$, a $f^\mu \in [0,1]$, 
equivalently,
a convex combination $\bar x(\mu)=\bar x^{f^{\mu}}$ of $\bar x^{sup}$ and
$\bar x^{inf}$, such that $\Delta_\mu (\bar x(\mu))$ converges to 
${q_\f^y}^*$. This gives the desired family 
of mean trajectories. 
We summarize the numerical scheme in Algorithm \ref{FPalgorithm}.
\begin{algorithm}[H]
\caption{Desirable near fixed point algorithm}
\label{FPalgorithm}
\begin{algorithmic}[1]
\State Initiate $n_2>n_1>1$, $t_0,d\mu,e_1,e_2,\gamma$
\Procedure{Generation of $\bar x^{sup}$ and $\bar x^{inf}$}{}
\State Apply $q_t^y=n_1 {q_\f^y}^*$ for  $t\in [0,t_0]$ and $q_t^y={q_\f^y}^*$ for $t>t_0$, and generate $\bar x^{sup} = \mathcal{T}(q_t^y)$. 
\State Apply $q_t^y= n_2{q_\f^y}^*$ for  $t\in [0,t_0]$ and $q_t^y={q_\f^y}^*$ for $t>t_0$, and generate $\bar x^{inf} = \mathcal{T}(q_t^y)$
\State $\mu_{sup}=\frac{{q_\f^y}^*}{\lim\limits_{t \to \f}\Delta_1(\bar x^{sup})}$ and 
$\mu_{inf}=\frac{{q_\f^y}^*}{\lim\limits_{t \to \f}\Delta_1(\bar x^{inf})}$
\EndProcedure
\State Initiate $\mu$ between $\mu_{sup}$ and $\mu_{inf}$
\Procedure{Gradient descent to compute $\mu_*$}{}
\While{ $err_1>e_1$ }
\For{$i=\mu$ and $i=\mu+d\mu$}
\State $f^m=0$, $f^M=1$, $\bar x^m = \bar x^{inf}$,
$\bar x^M = \bar x^{sup}$
\Procedure{Computation of $\bar x(i)$}{}
\While{$err_2>e_2$}
\State $f=(f^m+f^M)/2$ 
\State $\bar x(i)=f \bar x^{m}+(1-f)\bar x^{M}$
\If{$\lim\limits_{t \rightarrow \infty} \Delta_i(\bar x^f) \geq {q_\f^y}^*$}
\State $f^M=f$, $\bar x^M=\bar x(i)$
\Else
\State $f^m=f$, $\bar x^m=\bar x(i)$
\EndIf
\State $err_2=|\lim\limits_{t \rightarrow \infty} \Delta_i(\bar x(i)) - {q_\f^y}^*|$
\EndWhile
\EndProcedure
\EndFor
\State $\nabla=\big(\|\bar x(\mu+d\mu)-\mathcal{M}_{\mu+d\mu}\big(\bar x(\mu+d\mu)\big)\|_{L_2} -\|\bar x(\mu)-\mathcal{M}_{\mu}\big(\bar x(\mu)\big)\|_{L_2} \big)/d\mu$
\State $\mu=\mu-\gamma \nabla$
\EndWhile
\State $err_2=|G|$
\EndProcedure
\State Output: $\mu_*=\mu$ and $\bar x(\mu)$ a desirable near fixed point
\end{algorithmic}
\end{algorithm}

\section{SIMULATIONS}\label{sec:sim}

For our numerical experiments we simulate a 
population of 200 space heaters. We consider a 
uniform population of heaters; adopt a single
layer ETP model as given in \eqref{eqn:etp}, 
where the capacitance $(C_a)$ and 
conductance $(U_a)$ parameters \cite{param} are chosen 
to be 0.57 kWh/\degree\ and 0.27 kW/\degree\ 
respectively, the ambient temperature is 
set to -10 \degree, and the volatility 
parameter is set to 0.15 
$\degree/\sqrt{h}$. The initial 
temperatures of the heaters are drawn from 
a Gaussian distribution with a mean of 
21\degree\ and a variance of $1$. The cost 
function parameters $\delta$, $q^{x_0}$ and 
$r$ are uniformly chosen to be $0.001$, 
$200$ and $10$ respectively. We consider 
two scenarios. In the first scenario 
the heaters are controlled by the mean-field controllers \eqref{eqn:colOpt}, while 
in the second scenario they apply, for comparison purposes, the LQG 
controllers of 
Section \ref{sec:clasTrack}. We limit the size of computations by considering  a finite time horizon of length $T=3$ hours. For the mean field game based controller, we use the function $g(x)=\mu_* x$, where $\mu_*=1484$ is computed using the algorithm of Section \ref{num}. 

\begin{figure}[h]
\centering
\includegraphics[width=0.9\hsize,height=0.7\hsize]{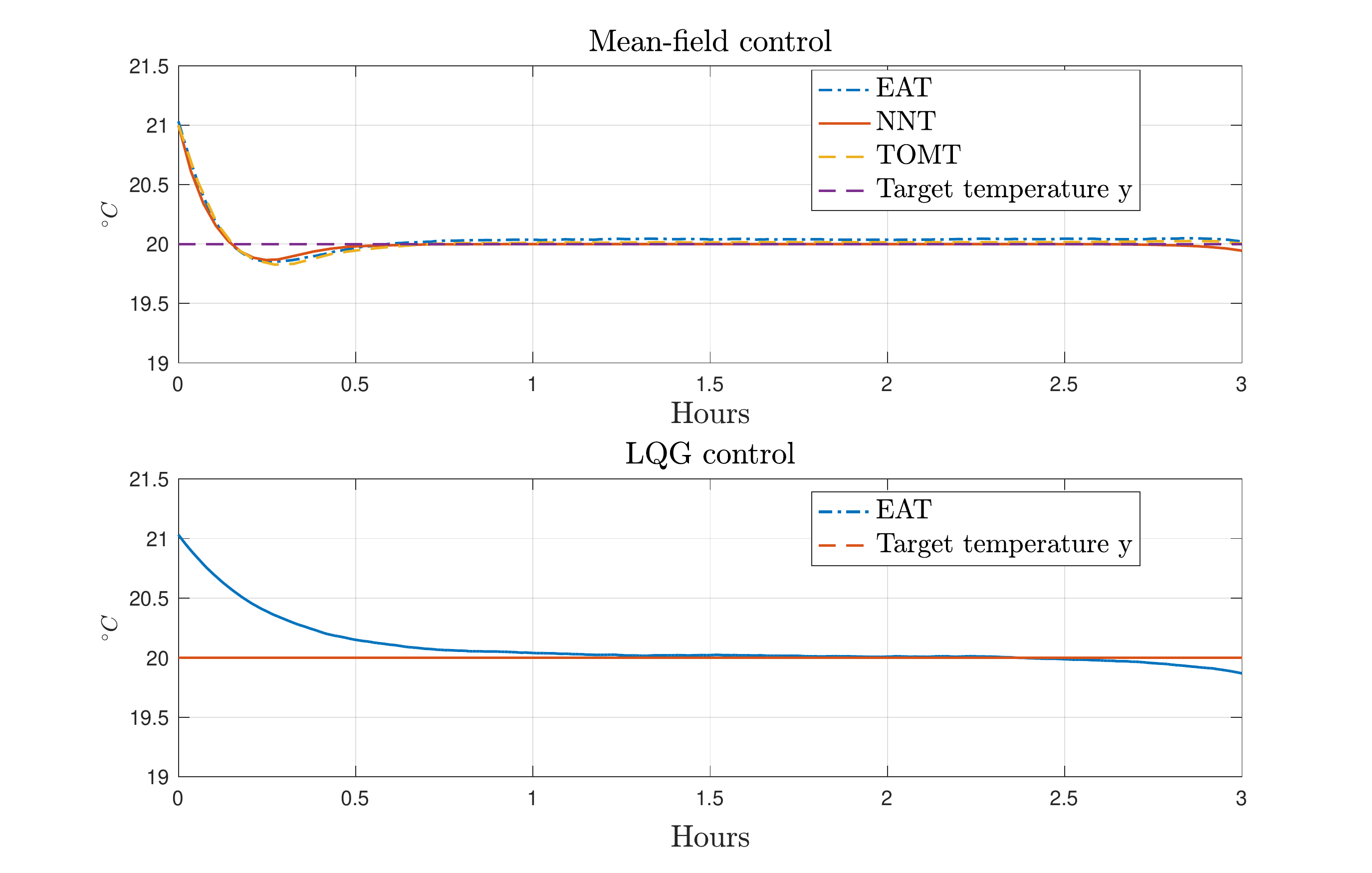}
\caption{NNT and EAT in the mean-field (linear $g$) and LQG control cases}
\label{fixedpoint}
\end{figure}

\begin{figure}[h]
\centering
\includegraphics[width=0.9\hsize,height=0.7\hsize]{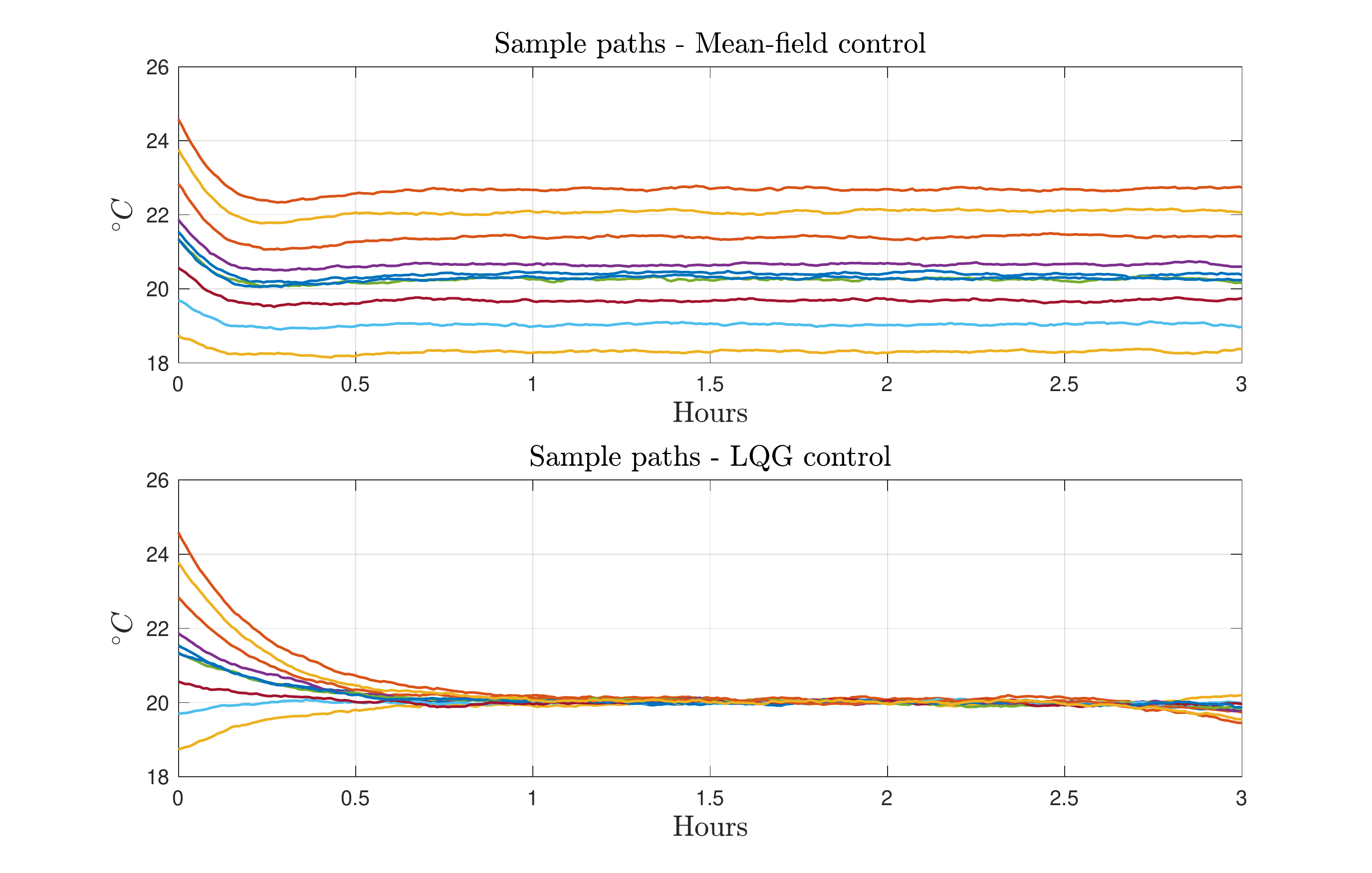}
\caption{Sample paths in the mean-field (linear $g$) and LQG control cases}
\label{samples}
\end{figure}

\rsa{Fig.} \ref{fixedpoint} shows the Near Nash Trajectory (NNT), i.e. near fixed point desirable temperature trajectory $\bar x$ generated by the algorithm of Section \ref{num}, the resulting Theoretical Output
Mean Trajectory (TOMT) $\mathcal{M}(\bar x)$   and the Empirical Average
Temperature (EAT) of the $200$ space heaters when they optimally respond to $\bar x$ by applying the mean-field controllers \eqref{eqn:colOpt}. As shown in the figure, the three trajectories are approximately identical and converge to the target 
temperature $y=20\degree$. 
This figure  illustrates also the EAT under the LQG controllers.
It should be noted that both approaches, MF and LQG, succeed in
forcing the empirical average space heater temperatures to converge to 
the target temperature. But, as shown in \rsa{Fig.}
\ref{samples} the user disturbance is much 
greater in the LQG case than it is in the mean field case. Indeed, this figure, which reports the temperature profiles of $10$ heaters when they apply respectively the mean field and LQG controllers, shows that all the 
temperatures must move to $y$ in the LQG
case, while in the mean field case the 
temperatures stay close to their 
initial values, and individual devices contribution is in direct relation with their initial energy content. 

\begin{figure}[h]
\centering
\includegraphics[width=0.9\hsize,height=0.7\hsize]{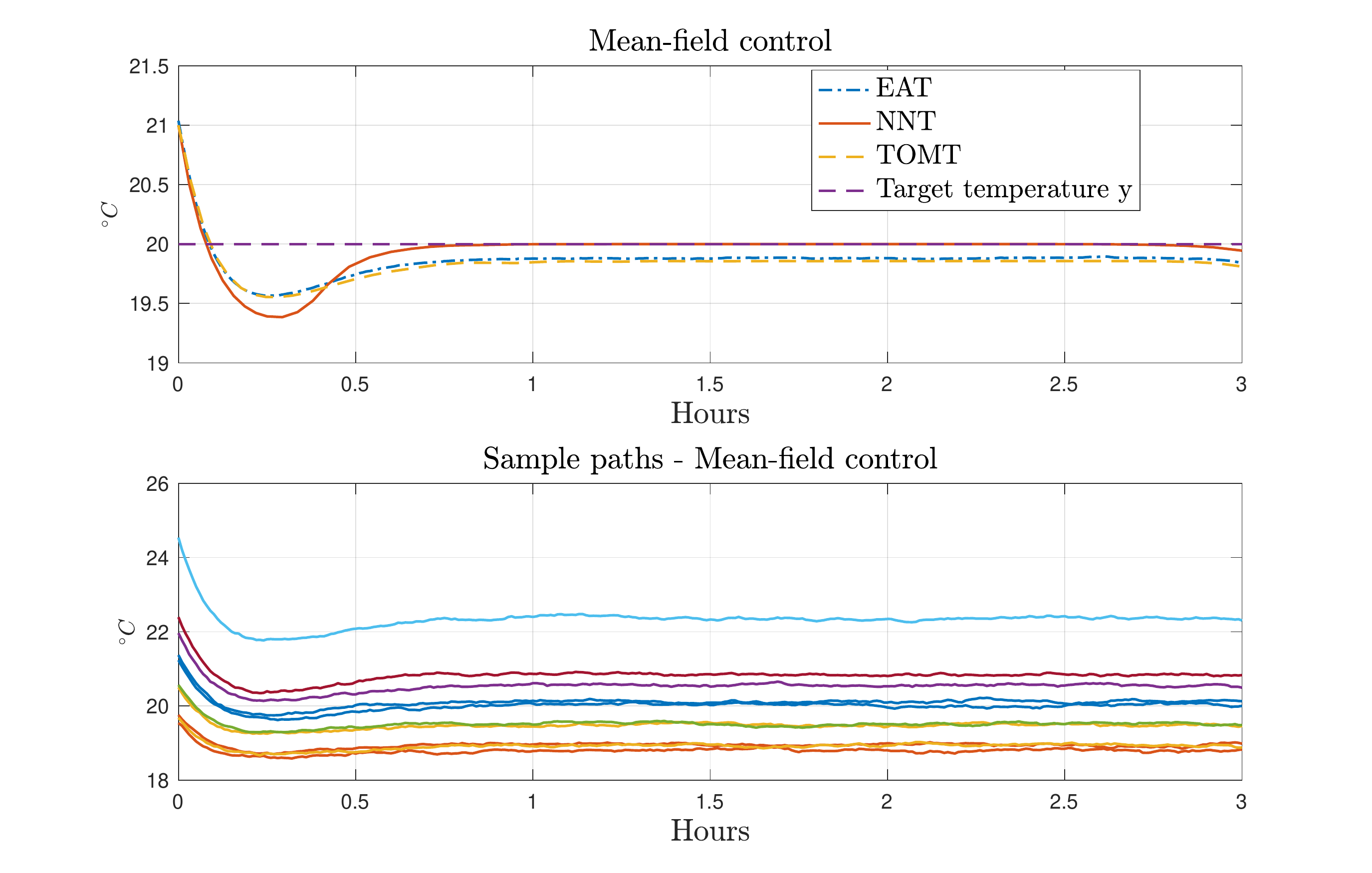}
\caption{NNT, EAT and sample paths in the mean-field (exponential $g$) case}
\label{meanexp}
\end{figure}

Next, we consider the case where $g(x)=\mu_*(e^{3x}-1)$, with $\mu_*=218$ is determined by the 
algorithm of Section \ref{num}. As shown in \rsa{Fig.} \ref{meanexp}, the TOMT reaches for the 
first time the 
target faster than the case with linear $g$, but needs 
more time to resettle on $y$. This is due to the fact 
that when $\bar x > y$, the dominant part of 
$g$ is the exponential. This creates a high pressure 
field towards $y$. When $\bar x$ undershoots $y$, the linear part of $g$, the scalar $-1$, becomes the dominant part, and the mean climbs slowly towards $y$.

\begin{figure}[h]
\centering
\includegraphics[width=0.9\hsize,height=0.7\hsize]{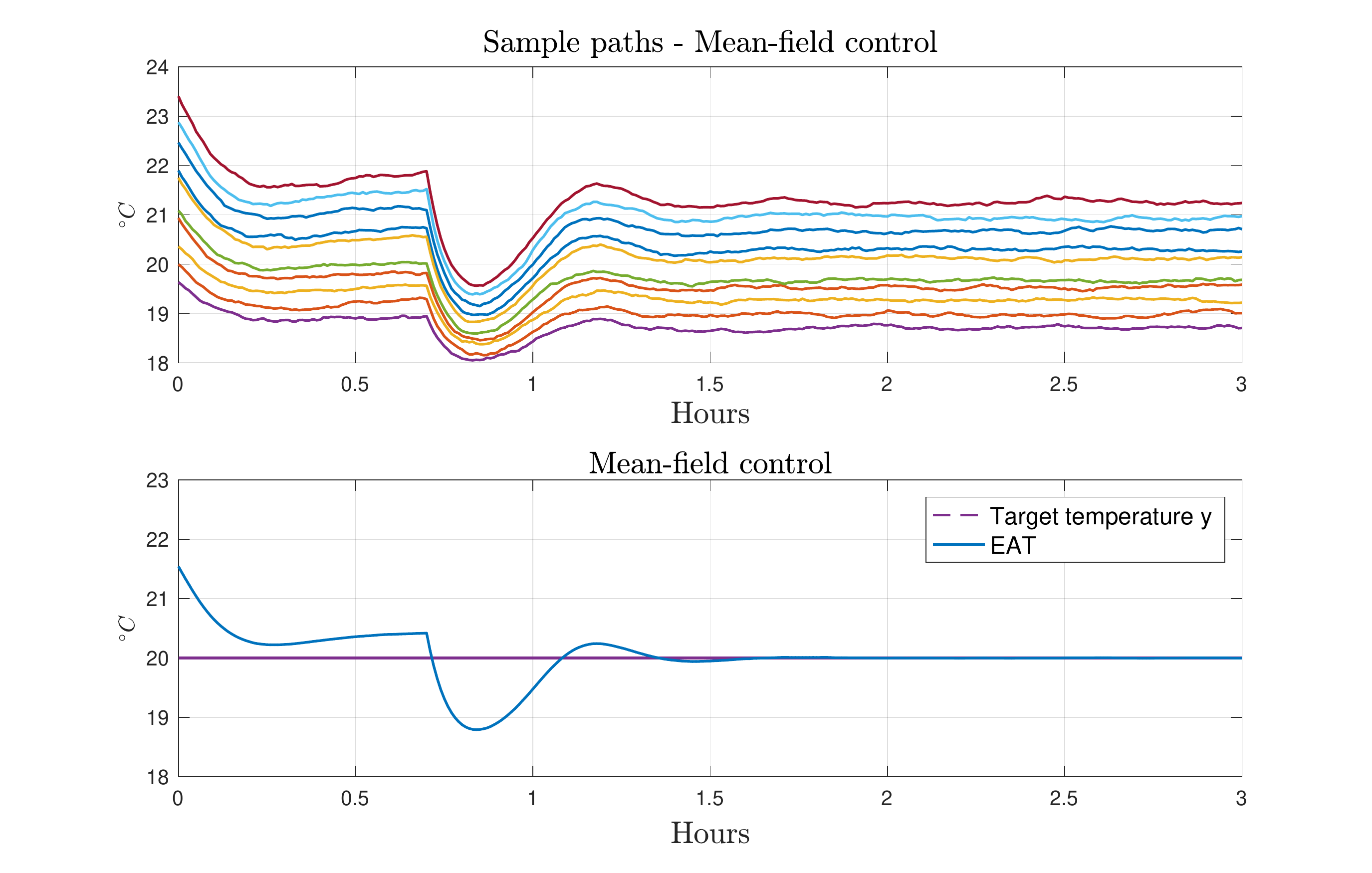}
\caption{Robustness of the mean field integral control}
\label{robust}
\end{figure}

Finally, we consider the case where the
heaters have biased a priori estimates of the initial mean and 
ambient temperatures. We assume that  the 
estimates are respectively
equal to $21\degree$ and $-10\degree$, while  the actual values are respectively 
equal to $21.5\degree$ and $-11\degree$. 
If the heaters apply the mean field controllers with linear $g$, which are computed based on the estimates, then 
the EAT will not converge to the target temperature $y=20\degree$. 
To compensate the estimation error, the heaters apply at the beginning the mean field controllers. When the EAT reaches a steady state, the controllers switch to the steady-state mean field controllers, i.e. \eqref{eqn:colOpt} where the coefficient $q_t^y$ is replaced by a constant equal to the current $q_t$ computed using the EAT. This constant is updated continuously, and 
the controller recomputed at each instant. The results 
are illustrated in \rsa{Fig.} \ref{robust}. The mean field controllers are 
applied between $t=0$ and $t=0.75$ hours.
Because of the biased mean and ambient temperatures estimates, the EAT reaches a steady-state value equal to $20.5\degree$ at $t=0.75$ hours.  At this instant, the heaters switch to the 
steady-state mean field controllers. As a 
result, the controllers compensate the 
estimation error and force the EAT to converge to $y=20\degree$.  


\section{CONCLUDING REMARKS}\label{sec:conc}

\rma{In this paper, the power of MFG based control methodologies has been applied to the problem of modulating the aggregate load of  vast numbers of energy storage associated devices in power systems, in particular electric space heaters (or air conditioners), acting as a buffer against the intermittency of renewable energy sources such as solar and wind, or as a virtual battery coordinated by an aggregator entity on the energy markets.  The MFG framework combines the ability of bottom up modeling approaches to answer ``what if'' types of questions arising in demand response or load dispatch schemes, with the decentralized control theoretic potential of game theory. From technically challenging in terms of complexity of control and monitoring needs,   the large numbers  involved  in the load management of residential and commercial loads, are turned  into an advantage as the laws of large numbers exploited in MFG formulations kick in. A novel integral control scheme acting on  the cost coefficients of a suitably engineered LQG individual agent cost is introduced, and a class of collective target following problems is defined.   The corresponding fixed point equations characterizing Nash equilibria are shown to have at least one solution. The latter may or may not be desirable (i.e. converging to target). An algorithm for a desirable near Nash equilibrium  is presented; agents contribute according to their current abilities, yet the mean dwelling temperature converges to the target.
The mean field based integral control formulation adds robustness to uncertainties  in the elemental device  models formulation , and are a first step towards an extension of the classical disturbance rejection control theory into collective dynamics control problems. In future work, we hope to develop MFG based collective dynamics formulations  whereby not only piecewise constant slowly changing perturbations can be rejected, but also disturbances within arbitrary signal classes. Furthermore, we intend to study  cooperative collective target tracking solutions, as well as the impact of saturation constraints on the synthesis of control laws.} 

\begin{ack}{Acknowledgment }        The authors gratefully acknowledge the support of Natural Resources Canada, and the Natural Sciences and Engineering  Research Council of Canada.  
\end{ack}


\appendix

\section{Preliminary results}\label{gLemmas}

\begin{pf*}{Proof of Lemma \ref{counterexample}}
The idea  for finding a counter example to continuity in the sup norm is 
to construct a sequence of mean trajectories
\rsa{$\{\bar{x}^{(n)}\}_n \in \mathcal{G}$}  (recall that $\mathcal{G}:= \{x: x \in \cb , \quad z \leq x(t) \leq x_0 \}$) that converges uniformly to $y$, in a way that the corresponding
coefficients \rsa{$q^{(n)}_t$} increase to 
infinity as $t \conv \infty$, for each $n \geq 0$. Based on Remark \ref{rem:optinterp} in Section \ref{cctmf}, we expect that $\lim\limits_{t \conv \f}\mathcal{M}(\bar x^{(n)})(t)=z$, for each $n$.  Moreover, $\mathcal{M}(y)(t) = \bar x_0$. Hence, $\|\mathcal{M}(\bar x^{(n)})- \mathcal{M}(y)\|_\infty$ does not converge to zero as $n \conv \infty $.
In particular, consider the sequence $\bar{x}^{(n)}_t=y+0.5^n$.
We have $\lim\limits_{n \rightarrow \infty}\|\bar{x}^{(n)}-y\|_\infty=0$.
The coefficients $q^y$ that correspond
to $\bar x^{(n)}_t$ and $y$ are
respectively $\Delta(\bar x^{(n)})=q_t^{(n)}=0.5^n t$ and $\Delta(y)=0$, 
and the corresponding Riccati equations 
are $(\pi_t^s)^{(n)}$ and $\pi_b^s$. 
Since $q_t^{(n)}=0.5^n t$ increases to 
infinity as $t \conv \infty$, one can show
that the corresponding Riccati solution $(\pi_t^s)^{(n)}$ also increases with $t$ to infinity.
We have $\mathcal{M}(y)(t)=\bar{x}_0$.
Define  $A^s_n(\tau)=a^s+\delta+(b^s)^2 (\pi_\tau^s)^{(n)} r^{-1}$ and
$f_n^s(\eta)=\int_\eta^\infty \exp\Big( -\int_\eta^\gamma A^s_n(\tau) d\tau \Big)d\gamma.$
$f_n^s$ is the unique bounded solution of 
$
df_n^s/d\eta = A^s_n(\eta)f_n^s -1.
$
Since $f_n^s$ is bounded, $(\pi_\eta^s)^{(n)}$ is increasing in $\eta$ and  $A^s_n(\eta)>0$, one can show that 
$
0\leq f_n^s(\eta) \leq 1/A^s_n(\eta).
$
Thus, $f_n^s$ is decreasing with time and it converges to zero as $t \conv \infty $. Now define the function
$g_n^s(t)=\int_0^t
\exp\Big(-\int_\eta^t\left(A^s_n(\tau)-\delta\right)d\tau\Big)  f_n^s(\eta) d\eta$. $g_n^s$ is the unique solution of 
$
dg_n^s/dt= -\left(A^s_n(t)-\delta\right)g_n^s(t)+f_n^s(t),
$
with $g_n^s(0)=0$.
$A^s_n(t)-\delta>0$, thus one can show that $\lim\limits_{t \rightarrow \infty} g_n^s(t)=0$. In 
view of \eqref{op_M}, we get
$\lim\limits_{t \rightarrow \infty} \T^s\circ \Delta(\bar{x}^{(n)})(t)=z$ for $1 \leq s \leq m$, which gives $\lim\limits_{t \rightarrow \infty} \M(\bar{x}^{(n)})(t)=z$. This implies 
that $\|\mathcal{M}(\bar{x}^{(n)})-\mathcal{M}(y)(t)\|_\infty \geq |\bar{x}_0-z|$, and 
$\mathcal{M}$ is not continuous on $\mathcal{G} \subset(\cb,\|.\|_\infty)$.
\end{pf*}
\begin{lem} \label{lem_1}
The following statements hold:
\begin{enumerate}[i.]
\item $(\ck,\|.\|_k)$ is a Banach space. 
\item $\mathcal{G}$ is a nonempty bounded closed convex subset of $(\ck,\|.\|_k)$.
\item If $\mathcal{F} \subset \mathcal{G}$ is a family of equicontinuous functions w.r.t. the sup-norm $\|\|_\f$, then $\bar{\mathcal{F}}$ is a compact subset of 
$(\ck,\|.\|_k)$.
\end{enumerate}
\end{lem}
\begin{pf}
i) $\ck$ is obviously a vector
space. Let $x^{(n)}$ be a Cauchy sequence in $\ck$. Fix a $t \geq 0$. We have for all 
$n,m>0$,
$
e^{-kt}|x^{(n)}_t-x^{(n+m)}_t| \leq \|x^{(n)}-x^{(n+m)}\|_k
$. Hence, $x^{(n)}_t$ is a Cauchy sequence in 
$\mathbb{R}$. We define $x$ to be the pointwise limit of $x^{(n)}$. Fix $t \geq 0$ and $\epsilon>0$. There exists $n_0(\epsilon)$, such 
for all $n>n_0$, $\|x^{(n)}-x^{(n+m)}\|_k<\epsilon$, for all $m>0$. Thus for all $n>n_0$ and $m>0$
\begin{align*}
e^{-kt}|x^{(n)}_t-x_t| &\leq 
\|x^{(n)}-x^{(n+m)}\|_k + e^{-kt}|x^{(n+m)}_t-x_t|\\ 
&\leq \epsilon + e^{-kt}|x^{(n+m)}_t-x_t|
\end{align*}
Hence, for all $t$, 
$
e^{-kt}|x^{(n)}_t-x_t| \leq  
\epsilon + \lim\limits_{m \conv \infty} e^{-kt}|x^{(n+m)}_t-x_t| = \epsilon.
$
Therefore, $\|x^{(n)}-x\|_k \leq \epsilon$, which 
implies that $x^{(n)}$ converges to $x$ under $\|\|_k$. It remains to show that $x \in \ck$. In fact, $x^{(n)}$ converges to $x$ under the norm
$\|\|_k$ means that the function $e^{-kt}x^{(n)}_t$
converges uniformly to $e^{-kt}x_t$. Thus,
$e^{-kt}x_t$ is continuous and bounded, and as a result 
$x$ is continuous and $\sup\limits_{t \in [0,\infty)} |e^{-kt}x_t| < \infty$. This implies that $x \in \ck$ and $\ck$ is a Banach space. 

ii) The proof of non-emptiness
and convexity are straightforward. For the 
boundedness, it suffices to remark that 
$|e^{-kt}x_t| \leq |x_t|$. It remains to 
prove that $\mathcal{G}$ is closed. Let $x^{(n)} \in \mathcal{G}$ a sequence that converges 
to $x$ in $\ck$. This implies that $x^{(n)}$ converges pointwise to $x$. Thus, for all
$t \geq 0$, $z \leq x_t \leq \bar{x}_0$. Hence, $x \in \mathcal{G}$ and $\mathcal{G}$ is closed.

iii) $\bar{\mathcal{F}}$ is a closed subset of the Banach space $\ck$. To prove the
compactness, it is sufficient to show that 
$\mathcal{F}$ is totally bounded \cite[Appendix A, Theorem A4]{1973Ru}, that is
for every $\epsilon>0$, $\mathcal{F}$ can be
covered by a finite number of balls of radius 
$\epsilon$. Let $\epsilon>0$, then, there
exists a time $T(\epsilon)>0$, such that 
$|e^{-kt}(\bar{x}_0-z)| < \epsilon$ for
all $t \geq T$. The set $\mathcal{F}_T=\{f_{|[0,T]}, \text{ for } f \in \mathcal{F}\}$
is uniformly bounded and equicontinuous in
$(\textbf{C}[0,T],\|\|_{[0,T]})$. Hence,
by Arzela-Ascoli Theorem \cite{1973Ru},
it is totally bounded
in $(\textbf{C}[0,T],\|\|_{[0,T]})$. Here, $\textbf{C}[0,T]$ denotes the set of continuous functions on $[0,T]$, $\|\|_{[0,T]}$ the corresponding sup-norm, and $f^{(j)}_{|[0,T]}$ denotes the restriction of $f^{(j)}$ to 
$[0,T]$.
Thus, there exist $f^{(1)}_{|[0,T]},\dots,f^{(n)}_{|[0,T]} \in 
\mathcal{F}_T$ such that $\mathcal{F}_T = \cup_{i=1}^n B(f^{(i)}_{|[0,T]},\epsilon)$, where the 
balls $B(f^{(i)}_{|[0,T]},\epsilon)$ are defined w.r.t. $(\b{C}([0,T],\|\|_{[0,T]})$. Let 
$f \in \mathcal{F}$, then there exists 
$1 \leq i \leq n$ such that $\sup\limits_{t \in [0,T]}|f(t)-f^{(i)}(t)| < \epsilon$. Hence,
$\sup\limits_{t \in [0,T]}e^{-kt}|f(t)-f^{(i)}(t)| < \epsilon$. Moreover, for all $t\geq T$, 
$e^{-kt}|f(t)-f^{(i)}(t)| \leq e^{-kt}|\bar{x}_0-z| < \epsilon$. Hence, $\|f-f^{(i)}\|_k<\epsilon$. Thus, $\mathcal{F}=\cup_{i=1}^n B(f^{(i)},\epsilon)$,
where the balls are now defined w.r.t.  
$(\ck,\|.\|_k)$.
This implies that $\mathcal{F}$ is totally bounded in $(\ck,\|.\|_k)$.
\end{pf}

\section{\rsa{Proof of the fixed point existence theorem}}\label{proofs}

Before showing the existence of a 
fixed point for $\mathcal{M}$, we need 
the following preliminary results \rsa{to apply Schauder's Theorem.}  

\begin{lem} \label{lem_2}
$Im(\mathcal{M}) \subset \mathcal{G}$.
\end{lem}
\begin{pf}
As noted in Remark \ref{rem:optinterp}, 
$\bar x^s = \T^s(q_t)$, $1 \leq s \leq m$,
is the optimal state of the optimal
control problem \eqref{optimalinterp}. We denote by $u_*^s$ the optimal control law of \eqref{optimalinterp}.
At first, we show by contradiction that $\bar{x}^s \leq \bar{x}_0$.
Suppose that there exists $t_0>0$ such 
$\bar{x}_{t_0}^s>\bar{x}_0$. Hence, there exist
$t_1<t_2$ such that $\bar{x}_{t_1}^s=\bar{x}_0$
and $\bar{x}_{t_2}^s=\bar{x}_0$ if $t_2<\infty$, and such that $\bar{x}_t^s>\bar{x}_0$ for all $t \in (t_1,t_2)$.
Let the control law $v$ be equal to $u_*^s$ on $[t_1,t_2]^c$, and to $0$ on $[t_1,t_2]$.
This means that the state 
that corresponds to $v$ is equal to $\bar x_0$ on $[t_1,t_2]$, and to the optimal
state $\bar x^s$ otherwise.
Hence, considering the cost in \eqref{optimalinterp},
$\inf\limits_u J(u) > J(v)$, which is impossible. 
We now show the $z$ lower bound inequality by contradiction. We assume without loss of 
generality that $\bar x_0=0$.
Suppose that there exists $t_0>0$ such 
$\bar{x}_{t_0}^s<z$. Hence, there exist
$t_1<t_2$ such that $\bar{x}_{t_1}^s=z$
and $\bar{x}_{t_2}^s=z$ if $t_2<\infty$, and such that $\bar{x}_{t}^s<z$ for all $t \in (t_1,t_2)$.
We define the control 
law $v$ to be equal to $u_*^s$ on $[t_1,t_2]^c$, 
and to $a^sz/b^s$ otherwise. Hence, the state that corresponds to $v$ is equal to $z$ on 
$[t_1,t_2]$, and to $\bar{x}^s$ on $[t_1,t_2]^c$. We have, 
\begin{align*}
I_1&=\int_{t_1}^{t_2}e^{-\delta t}v^2dt=
\int_{t_1}^{t_2}e^{-\delta t}\frac{(a^s)^2}{(b^s)^2}z^2dt\\
I_2&=\int_{t_1}^{t_2}e^{-\delta t}(u_*^s)^2dt=
\int_{t_1}^{t_2}e^{-\delta t}\Big\{\frac{(a^s)^2}{(b^s)^2}(\bar{x}^s)^2\\
&+\frac{1}{(b^s)^2}\left(\frac{d\bar{x}^s}{dt}\right)^2+\frac{2a^s}{(b^s)^2}\bar{x}^s\frac{d\bar{x}^s}{dt}\Big\}dt \\
&>I_1 +\frac{a^s}{(b^s)^2}\int_{t_1}^{t_2}e^{-\delta t}2\bar{x}^s\frac{d\bar{x}^s}{dt}dt.
\end{align*}
The inequality follows from the fact that
$\bar x_t^s< z$ on $[t_1,t_2]$ with $z<0$ according to our assumption.
Furthermore, using the integration by parts we get that
\begin{align*}
&\int_{t_1}^{t_2}e^{-\delta t}2\bar{x}^s\frac{d\bar{x}^s}{dt}dt= z^2(e^{-\delta t_2}-e^{-\delta t_1})+\int_{t_1}^{t_2}\delta e^{-\delta t}(\bar{x}^s)^2dt \\
&\geq z^2(e^{-\delta t_2}-e^{-\delta t_1})+
\int_{t_1}^{t_2}\delta e^{-\delta t}z^2dt =0.
\end{align*}
Hence, $\inf\limits_u J(u)=J(u_*^s) > J(v)$, which is impossible. As a result $\xf^s \in \G$, for $1 \leq s \leq m$, which implies 
that $\sum_{s=1}^m n_s\bar x^s \in \G$, and  
proves the result.
\end{pf}

\begin{lem} \label{lem_3}
Under \ass{liptass},  the following statements hold:
\begin{enumerate} [i.]
\item There exists $k_0>0$, such that
$\forall \bar{x} \in \mathcal{G}$, 
$\Delta(\bar x)(t)=q_t^y \leq k_0 t$.
\item There exist $k_1>0$ and $k_2>0$, such that $\forall \bar{x} \in \mathcal{G}$, 
$\pi_t^s \leq k_1 t+ k_2$. Here $\pi_t^s$ is the 
solution of the Riccati equation \eqref{eqn:piwatMFgen}, with $q^y=\Delta(\bar x )$.
\item Fix $0<k<2\min\limits_s a^s+\delta$. There exists $k_3>0$, such that $\forall \bar x,\bar x'\in \mathcal{G}$,
we have $\|\pi^s-(\pi^s)'\|_k \leq k_3\|\bar x-\bar x'\|_k$. Here $\pi_t^s$ and  (resp. $(\pi^s)'$) is the 
solution of the Riccati equation \eqref{eqn:piwatMFgen}, with $q^y=q:=\Delta(\bar x )$ (resp. $q^y=q':=\Delta(\bar x' )$).
\end{enumerate}
\end{lem}
\begin{pf}
i) We have $\forall \bar x \in \mathcal{G}$,
$
q_t^y= \left |  \int_0^t g(\bar{x}_\tau-y_\tau)d\tau \right | \leq \max\limits_{|x| \leq |\bar{x}_0-z|}|g(x)|t \triangleq k_0 t.
$

ii) $\forall \bar x \in \mathcal{G}$,
$d\pi_t^s/dt= (2a^s + \delta +(b^s)^2 r^{-1}\pi_t^s) \pi_t^s 
-q_t^y - q^{x_0}.
$
Hence,
\begin{align*}
\pi_t^s &= \int_t^\infty \exp \left ( 
-\int_t^\eta (2a^s+\delta+ \frac{(b^s)^2}{r} \pi_\tau^s)
d\tau\right)(q_\eta^y + q^{x_0})  d\eta \\
&\leq 
\int_t^\infty (k_0\eta+q^{x_0}) e^{-2a^s(\eta-t)} d\eta\\
&\leq \frac{k_0}{2\min\limits_s a^s}t + \frac{k_0}{4\min\limits_s (a^s)^2}+ \frac{q^{x_0}}{2\min\limits_s a^s} \triangleq k_1t + k_2.
\end{align*}

iii) We have  $\forall \bar x,\bar x' \in \mathcal{G}$,
\begin{align*}
&e^{-kt}|q_t-q_t'| \leq e^{-kt} 
\int_0^t \left |g(\bar{x}_\tau-y_\tau)-g(\bar{x}_\tau'-y_\tau) \right | d\tau \\
&\leq
\lambda e^{-kt} 
\int_0^t e^{k\tau}e^{-k\tau}|\bar{x}_\tau-\bar{x}'_\tau| d\tau \leq \frac{\lambda}{k} \|\bar{x}-\bar{x}'\|_k,
\end{align*}
where $\lambda$ is the Lipschitz constant of 
$g$ on the compact subset $[-|\bar{x}_0-z|,|\bar{x}_0-z|]$.
We define $\bar{\pi}_t^s=e^{-kt}\pi_t^s$, 
$(\bar{\pi}_t^s)'=e^{-kt}(\pi_t^s)'$, $\tilde{\pi}_t^s=\bar{\pi}_t^s-(\bar{\pi}_t^s)'$, $\tilde{q}_t=e^{-kt}(q_t-q_t')$ and $\tilde{A}^s_t=2a^s + \delta -k+e^{kt}\frac{(b^s)^2}{r}(\bar{\pi}_t^s+(\bar{\pi}_t^s)')$.
We obtain that,
$
d\tilde{\pi}_t^s/dt= \tilde{A}^s_t\tilde{\pi}_t^s 
-\tilde{q}_t.
$
Therefore,
\begin{align*}
|\tilde{\pi}_t^s| &= \Big|\int_t^\infty \tilde{q}_\eta \exp \Big ( 
-\int_t^\eta \tilde{A}^s_\tau
d\tau\Big)  d\eta\Big|  \\
&\leq \frac{\lambda}{k(2\min\limits_s a^s+\delta-k)}\|\bar{x}-\bar{x}'\|_k \triangleq k_3 \|\bar{x}-\bar{x}'\|_k.
\end{align*}
\end{pf}

\begin{lem}  \label{lem_4}
$\overline{\mathcal{M}(\mathcal{G})}$  is compact in 
$(\ck,\|.\|_k)$.
\end{lem}
\begin{pf}
Based on the third point of Lemma \ref{lem_1}, it suffices to show that 
$\mathcal{M}(\mathcal{G})$ forms a family of equicontinuous functions w.r.t. the sup-norm. This can be done 
by showing that the family of derivatives of the functions in  $\mathcal{M}(\mathcal{G})$ is uniformly 
bounded.
We have for all $\bar{x} \in \mathcal{G}$,
\begin{align*} 
&\frac{d\T^s\circ\Delta(\bar{x})}{dt}(t)=A_t^s\phi^s(t,0)(\bar{x}_0-z) +\\
&C^s \int_t^\infty \psi^s(\tau,t)d\tau d\eta-C^s \int_0^t
A_t^s \\
& \times\phi^s(t,\eta)\int_\eta^\infty \psi^s(\tau,\eta)d\tau d\eta
\triangleq  \xi_1 + \xi_2
+ \xi_3. \nonumber
\end{align*}
where $A_t^s=a^s+(b^s)^2r^{-1} \pi_t^s$ and $C^s$ is defined in \eqref{op_M}.
The function $\phi^s(t,0)$ is the unique solution 
of the following differential equation,
$
d\phi^s(t,0)/dt=-A_t^s\phi^s(t,0)
$
with $\phi^s(0,0)=1$. We show in the following that the 
product $\phi^s(t,0)A_t^s$ is bounded. But, $\phi^s(t,0)A_t^s$ satisfies 
$
d(\phi^s(t,0)A_t^s)/dt=-A_t^s(\phi^s(t,0)A_t^s)+\dot{A}_t^s\phi^s(t,0).
$
Hence,
\[
\phi^s(t,0)A_t^s=A_0^se^{-\int_0^t A_\tau^s d\tau} + \int_0^t e^{-\int_\eta^t A_\tau^s d\tau}
\dot{A}_\eta^s\phi^s(\eta,0)d\eta.
\]
We have $|A_0^s| \leq a^s+(b^s)^2r^{-1}k_2$ and  $\|\phi^s(t,0)\|_\infty \leq 1$, where $k_2$ is defined in the second point of Lemma \ref{lem_3}. 
Fix $h \in \mathbb{R}$, and define 
the functions
$\tilde{\pi}^s_t=\pi^s_{t+h}-\pi_t^s$ and
$\tilde{q}_t=q_{t+h}^y-q_t^y$.
We have 
\begin{align*}
-\frac{d\tilde{\pi}_t^s}{dt}&=-\left(2a^s+\delta 
 +\frac{(b^s)^2}{r}(\pi_t^s+\pi_{t+h}^s) \right) \tilde{\pi}_t^s + \tilde{q}_t. 
\end{align*}
Hence,
\[
\tilde{\pi}_t^s=\int_t^\infty  \tilde{q}_\tau \exp \Big(-\int_t^\eta (2a^s+\delta 
 +\frac{(b^s)^2}{r}(\pi_\eta^s+\pi_{\eta+h}^s) )d \eta \Big)d\tau. 
\]
Thus, $|\tilde{\pi}_t^s|\leq h \max\limits_{|x| \leq |\bar{x}_0-z| }|g(x)|/2a^s$, which implies that 
$\lim\limits_{h \rightarrow 0} \left | \tilde{\pi}^s_t/h \right| \leq \max\limits_{|x| \leq |\bar{x}_0-z| }|g(x)|/2a^s$ and $\left \|d \pi^s_t/dt \right\|_\infty \leq \max\limits_{|x| \leq |\bar{x}_0-z| }|g(x)|/2a^s$.
Hence recalling the definition of $A^s$ above, $\dot{A}^s$ is uniformly bounded, and as
a result $\phi^s(t,0)A^s_t$ is uniformly bounded by $M^s>0$. This implies that $|\xi_1| \leq M^s |\bar{x}_0-z|\triangleq K_1^s$. Recalling the definition of $\psi^s(\tau,t)$ in Section \ref{failurecont} above, the second and
third terms satisfy the following inequalities,
\begin{equation*}
\begin{aligned} 
|\xi_2 |& \leq  
C^s \int_0^\infty e^{-a^s\tau} d\eta \triangleq K_2^s\\
 |\xi_3 |& \leq \frac{C^s}{a^s} \int_0^t
 A_t^s
\phi^s(t,\eta) d\eta,
\end{aligned}
\end{equation*}
We define on $[0,\infty)$, 
$
 f^s_t=\int_0^t
\phi^s(t,\eta) d\eta.
$
$f^s_t$ is the unique solution 
of 
$
df^s/dt=-A_t^sf^s_t + 1$, with $f_0^s=0$. Therefore,
$
d(f^sA^s-1)/dt=-A^s_t(f^s_tA^s_t-1)+\dot{A}_t^sf_t^s.
$
Solving this equation  gives
\[
f_t^sA_t^s-1=-e^{-\int_0^t A_\tau^s d\tau} + \int_0^t e^{-\int_\eta^t A_\tau^s d\tau}
\dot{A}_\eta^sf_\eta^sd\eta.
\]
We have $\|f_t^s\|_\infty \leq 1/a^s$. 
Hence, $f_t^sA_t^s-1$ is uniformly bounded. Denote $M_1^s$ the bound of $f_t^sA_t^s$. We get that 
the third term 
$|\xi_3 | \leq C^s M_1^s/a^s
 \triangleq K_3^s.
$
Finally, $\left\|d\mathcal{M}(\bar{x})/dt\right\|_\infty \leq \sum_{s=1}^m n_s(K_1^s + K_2^s+K_3^s)$, where $K_1^s$, $K_2^s$ and $K_3^s$ do not 
 depend on $q_t^y$. This proves the result.
\end{pf}

\begin{pf*}{Proof of Theorem \ref{thmFixed}}
The second point follows from Lemmas \ref{lem_1}, \rsa{\ref{lem_2}} \ref{lem_3},
\ref{lem_4} (\rsa{Lemma \ref{lem_2} guaranties that $\mathcal{M}$ stabilizes $\mathcal{G}$}), Schauder's fixed point Theorem
\cite{1973Ru}, and the first point 
of this theorem, which we show in the
remainder of this proof. Let $\bar x'$ and $\bar x''$ belong to $\mathcal{G}$. We denote respectively by $(\pi^s)'$ and $(\pi^s)''$ the 
corresponding Riccati solutions. We have,
\begin{align*}
&\T^s\circ \Delta(\bar{x}')(t)-\T^s\circ \Delta (\bar{x}'')(t)=\\
&\left(\phi^s_1(t,0)-\phi^s_2(t,0)\right)(\bar{x}_0-z)  \\
&+C^s \int_0^t
\phi_1^s(t,\eta)  \int_\eta^\infty (\psi_1^s(\tau,\eta)-\psi_2^s(\tau,\eta))d\tau d\eta\\
&+C^s \int_0^t
(\phi_1^s(t,\eta)-\phi_2^s(t,\eta))  \int_\eta^\infty \psi_2^s(\tau,\eta)d\tau d\eta\\
& \triangleq \xi_1 + \xi_2 + \xi_3,
\end{align*}
where $(\phi_1^s,\psi_1^s)$, and $(\phi_2^s,\psi_2^s)$  are the $(\phi^s,\psi^s)$ functions defined in \eqref{phidef}, and that correspond respectively to $(\pi^s)'$ and $(\pi^s)''$. $C^s$ is defined in \eqref{op_M}.
Using the inequality, $|e^{-x}-e^{-y}| \leq e^{-\min (x,y)}|x-y|$, for $x$ and $y$ positive, we obtain that 
\begin{align*}
|\xi_1|&\leq |\bar{x}_0-z|(b^s)^2r^{-1}e^{-a^st}\|(\pi^s)'-(\pi^s)''\|_k 
\int_0^te^{k\tau} d\tau \\
&\leq \lambda\frac{|\bar{x}_0-z|(b^s)^2}{rk^2(2\min_sa^s+\delta-k)}\|\bar x'-\bar x''\|_k e^{kt},
\end{align*}
where the second inequality follows from 
the third point of Lemma \ref{lem_3} 
and the definition of the constant $k_3$ as developed in Appendix \ref{proofs}.
Hence,
\[
\|\xi_1\|_k \leq \lambda\frac{|\bar{x}_0-z|(b^s)^2\|\bar x'-\bar x''\|_k}{rk^2(2\min\limits_s a^s+\delta-k)} \teq \lambda c_1^{sk} \|\bar x'-\bar x''\|_k.
\]
Using the same inequality as above, we 
get that
\begin{align*}
&|\xi_2| \leq \lambda \frac{(b^s)^2C^s}{rk^2(2\min\limits_sa^s+\delta - k)}\|\bar x'- \bar x''\|_k \times\\
&\int_0^t
e^{-a^s(t-\eta)}  \int_\eta^\infty e^{-(a^s+\delta)(\tau-\eta)}(e^{k\tau}-e^{k\eta})d\tau d\eta\\
& =\lambda\frac{L^s\|\bar x'- \bar x''\|_k (e^{kt}-e^{-a^st})}{(a^s+\delta-k)(a^s+k)},
\end{align*}
where $L^s=(b^s)^2C^s/rk(2\min\limits_sa^s+\delta - k)(a^s+\delta) $.
Therefore,
\begin{align*}
&\|\xi_2\|_k \leq \lambda\frac{L^s\|\bar x'- \bar x''\|_k}{(a^s+\delta-k)(a^s+k)} \teq \lambda c_2^{sk}\|\bar x'- \bar x''\|_k.
\end{align*}
We have that,
\begin{align*}
|\xi_3| &\leq  \lambda L^s\|\bar x'-\bar x''\|_k\int_0^t 
e^{-a^s(t-\eta)}(e^{kt}-e^{k\eta})d\eta\\
& \leq \lambda\frac{L^s}{a^s}\|\bar x'-\bar x''\|_k e^{kt}.
\end{align*}
Hence,
$\|\xi_3\|_k 
 \leq \lambda L^s (a^s)^{-1}\|\bar x'-\bar x''\|_k \teq \lambda c_3^{sk}\|\bar x'-\bar x''\|_k. 
$
Finally, $
R_k=\sum_{s=1}^m n_s (c_1^{sk}+c_2^{sk}+c_3^{sk}).
$
\end{pf*}



\bibliographystyle{plain}        
\bibliography{autosam} 
\end{document}